\documentclass[12pt,preprint]{aastex}

\begin{document}
\def\etal{{\it et al.\/}}
\def\cf{{\it cf.\/}}
\def\ie{{\it i.e.\/}}
\def\eg{{\it e.g.\/}}

\title{On the stability of shocks with particle pressure}
\author{{\bf Stefano Finazzi\footnote{Currently at SISSA, Trieste.}$\;\;$ and Mario Vietri}}
\affil{Scuola Normale Superiore, Pisa} {}
\begin{abstract}
We perform a linear stability analysis for corrugations of a
Newtonian shock, with particle pressure included, for an arbitrary
diffusion coefficient. We study first the dispersion relation for
homogeneous media, showing that, besides the conventional pressure
waves and entropy/vorticity disturbances, two new perturbation modes
exist, dominated by the particles' pressure and damped
by diffusion. We show that, due to particle diffusion into the 
upstream region, the fluid will be perturbed also upstream: we 
treat these perturbation in the short wavelength (WKBJ) regime. 
We then show how to construct a corrugational mode for
the shock itself, one, that is, where the shock executes free
oscillations (possibly damped or growing) and sheds perturbations
away from itself: this global mode requires the new modes. Then, using
the perturbed Rankine--Hugoniot conditions, we show that this leads
to the determination of the corrugational eigenfrequency. We solve
numerically the equations for the eigenfrequency in the WKBJ regime
for the models of \citeauthor{amatoblasi2005}, showing that they are stable. We
then discuss the differences between our treatment and previous work.
\end{abstract}


\section{Introduction}\label{sec:intro}

One of the major uncertainties still surrounding particle
acceleration around shocks is the level at which this saturates:
\ie, the total energy content in non-thermal particles, per unit
volume, $\cal{R}$, in units of the incoming fluid kinetic energy
density. This saturation level plays a very important role, for
instance, in discussions of the origin of cosmic rays as observed at
Earth: it has to be rather large ($\ga 0.1$) to allow SNe to provide
the observed flux. Also, discussions on the origin of UHECRs are
influenced by similar considerations: the idea that UHECRs are
accelerated by GRBs \citep{vietri1995,waxman1995} draws some measure of
support by the coincidence between the energy release rate of GRBs
in $\gamma$-ray photons and that required to account for
observations at Earth \citep{waxman2002,vietrial2003}.

It is certainly possible that this level of saturation is somehow
connected with the very uncertain particle injection mechanism,
and that saturation occurs, in many astrophysically important
situations, at very low levels. Yet, given the arguments
concerning SNe and GRBs mentioned above, it appears that at least
occasionally saturation must occur at rather large levels. This is
especially so in the case of Galactic cosmic rays: if in fact we
were to discard SNe as sources of cosmic rays, all other possible
classes of sources, being both less numerous and less energetic,
would force us to require saturation at even larger
levels\footnote{The situation is much less clear for UHECRs,
because many of the proposed classes of sources are purely
hypothetical, so that there are no constraints on their
properties.}.

A distinct possibility is that the saturation level is determined
not by the injection mechanisms, but by modifications which the
particles' pressure induces in the shock properties. After all, in
the test particle limit the particles' spectrum is ultraviolet
divergent (though marginally so, of course), and one may hope that,
since the particles' back-reaction on the fluid makes acceleration
less efficient (by reducing the velocity jump around the gas
sub-shock), a convergent spectrum will be obtained, with a definite
value for the parameter ${\cal R}$. After \citeauthor{malkov1997}'s seminal papers
\citep{malkov1997,malkoval2000,malkovdrury2001},
fully self-consistent solutions with particles' pressure properly
included have been obtained by \citet{amatoblasi2005}. In these
models, it appears that particles can carry away an arbitrarily
large fraction of the incoming energy flux, for sufficiently large
Mach number at upstream infinity. Even more intriguing is the fact
that these solutions have particles' spectra which are more, not
less, ultraviolet divergent than those in the test particle limit.
In fact, this divergence is arrested only by arbitrarily limiting
the largest individual energy to some fiducial value \citep{amatoblasi2005}.

These solutions then seem to beckon for a stability analysis, in the
hope that they are found to be unstable once ${\cal R}$ exceeds some
critical value. The instability we envision is of course the
corrugational instability for shocks, whereby ripples on the shock
surface become of larger and larger amplitude\footnote{In all of
this paper, the term corrugation instability is taken to include
also the spontaneous emission of sound waves by the shock.}. If this
instability were to exist, we might easily imagine that the shock is
substituted by a region of strong fluid turbulence, where particles
moving a few Larmor radii perceive only a small velocity difference
at the two ends of their free wandering, and thus acceleration to
high energies is made somewhat less efficient.

However, it is well-known \citep{landaulifshitz1987} that shocks in
polytropic fluids are stable against this kind of perturbations, and
against the spontaneous emission of sound waves as well. The hope
for the existence of an instability is based on a rather more subtle
argument. When the shock surface flaps, it sheds in the downstream
region pressure waves as well as entropy and vorticity
perturbations. We will show later that the last two do not couple to
particle perturbations, but pressure waves do, thus generating small
perturbations in the particles' distribution function, $\delta\!f$.
These particles will however return upstream by means of diffusion,
and here generate, by their sheer perturbed pressure, some more
fluid perturbations. Thus, not all energy shed by the shock is lost
forever toward upstream infinity, as is the case in the purely
hydrodynamical case, but some fraction of it returns to the shock to
generate more flapping. In fact, since pressure waves are strongly
damped by diffusion, and nearly all particles return to the upstream
section since the shock is Newtonian, we may conjecture that most
energy shed by the shock flapping makes it back to feed more
flapping, even though account must be taken of diffusion and phase
mixing. Is it possible that this sets up some strong reinforcement,
making the whole system unstable? This is the question we address in
this paper.

This question has of course been studied before, under somewhat
different conditions.  The diffusion coefficient of non-thermal
particles by the fluid ($\equiv D$) has been taken (by other
authors) as independent of the particle impulse: this is
conventionally referred to as the two-fluid approximation, because
the Boltzmann equation for the non-thermal particles can then
easily be recast into an equation for their pressure, thusly erasing
all references to the underlying distribution function. Besides
making the two-fluid approximation, \citet{monddrury1998} also
neglected diffusion altogether. A more complex analysis has been
presented by \citet{toptygin1999}, who included energy transport and
particle injection at the shock, but still in the two-fluid
approximation. In the following, we shall make no such
approximations: we will consider a finite diffusion coefficient $D$,
and it will be allowed to be an arbitrary function of both $p$, the
particle momentum, and of $\rho$, the local fluid density, so as to
at least mimic the increase in magnetic field strength due to flux
freezing. The arbitrary nature of the dependence of $D$ on $p$
automatically prevents the use of the two-fluid approximation, and
forces us to use the full Boltzmann-Skilling equation.

A word of caution is in order about some assumptions. We will
neglect all energy in the form of magnetic field and Alfv\'en waves;
of course this is necessary because the zero-th order solutions of
Blasi and collaborators do not include these effects, but in our
case this neglect requires one extra assumption, {\it i.e.} that the
time scales for energy to accumulate into any of these energy sinks,
$T_{in}$, and to flow out of each of them, $T_{out}$, be ordered
like this: $T_{out} \lesssim T_{in}$. If this inequality were
severely violated, the magnetic field or Alfv\'en waves might
acquire a significant fraction of the total energy, despite their
negligibility in the zero-th order solution, and make our treatment
completely irrelevant.

{}

In the following, we shall follow closely the treatment of the shock
corrugational instability given by \citet{landaulifshitz1987}. In
particular, we shall consider a steady-state shock in its own
frame, located at $z=0$, with fluid coming from the left, and
exiting from the right, so that all speeds are $>0$. Exactly like
\citeauthor{landaulifshitz1987}, we shall consider perturbations generated by
the shock flapping, so that there can be no incoming waves, from
either upstream or downstream infinity. In Section \ref{sec:homog}, we shall
consider perturbations in a homogeneous medium; this is perhaps a
tad boring, but it contains a number of new results which are of the
utmost importance later on. In Section \ref{sec:pertup} we briefly discuss
perturbations upstream, {\it i.e.} where the flow is inhomogeneous;
we show here that we can easily obtain the perturbations in the WKBJ
limit $k_y\rightarrow \infty$, which restricts our analysis to the 
regime $\lambda \lesssim L$, where $\lambda$ is the perturbation 
wavelength perpendicular to the shock, and $L$ is the typical size of
the region of inhomogeneity upstream. In Section \ref{sec:fitboundary} we discuss the 
boundary conditions on the perturbed particle distribution function,
and derive how it relates to the amplitude of the modes to which
particle perturbations couple. We present in Section \ref{sec:fluifcond} the
perturbed Rankine--Hugoniot conditions and in Section \ref{sec:eigenf} what fixes
the global corrugation mode eigenfrequency. In Section \ref{sec:results} we present
our numerical computations for the stability of the exact solutions
of the zero-th order problem by \cite{amatoblasi2005}. In Section
\ref{sec:discussion}, we will compare our results with other works in the literature,
and briefly summarize our work.

\section{Perturbations in a homogeneous medium}\label{sec:homog}

We give here, for future reference, our basic equations. They are
the conventional hydrodynamic equations:
\begin{equation}
\label{masscons}
\frac{\partial \rho}{\partial t} +
\bigtriangledown\cdot(\rho\vec{v}) = 0
\end{equation}
\begin{equation}
\label{momentumcons} \frac{\partial}{\partial t}\vec{v} +
\vec{v}\cdot\bigtriangledown \vec{v} =
-\frac{1}{\rho}\bigtriangledown (P+P_c)
\end{equation}
\begin{equation}
\label{entropycons} \frac{\partial s}{\partial t} + \vec{v}\cdot
\bigtriangledown s = 0
\end{equation}
which contain a term for the momentum exchange between the fluid
and the non-thermal particles represented by the gradient of the
particle pressure $P_c$, plus the usual Boltzmann equation in
\citet{skilling1975}:
\begin{equation}\label{boltzmann}
\frac{\partial f}{\partial t} =
\bigtriangledown\cdot(D\bigtriangledown f) -
\vec{v}\cdot\bigtriangledown f + \frac{1}{3}
\bigtriangledown\cdot\vec{v}\; p \frac{\partial f}{\partial p}\;.
\end{equation}
We assume $D = D(p,\rho)$ to be a given function of $\rho$ and $p$.

We consider small-amplitude perturbations around a homogeneous
solution where the particles are supposed to exert a
non-negligible pressure. First, we consider entropy
perturbations. Perturbations can be taken in the form
\begin{equation}
\delta\!s \propto \exp\left(\imath\omega t - \imath\vec{k}
\cdot\vec{r}\right)
\end{equation}
so to obtain, from the equation of entropy conservation, eq.
\ref{entropycons},
\begin{equation}
(\omega-uk_x)\delta\!s = 0\;\;\; \rightarrow \;\;\;\omega-uk_x =
0\;,
\end{equation}
where $u$ is the unperturbed fluid velocity in the x-direction.
Perturbation of the mass conservation equation yields
\begin{equation}
(\omega-u k_x)\delta\!\rho -\rho \vec{k} \cdot\delta\!\vec{v} = 0
\;\;\; \rightarrow \;\;\;\vec{k} \cdot\delta\!\vec{v} = 0\;.
\end{equation}
Perturbation of the momentum equation yields
\begin{equation}
(uk_x-\omega)\delta\!\vec{v} + \frac{\vec{k}}{\rho}(\delta\!P +
\delta\!P_c) = 0\;\;\; \rightarrow\;\;\; \delta\!P =
-\delta\!P_c\;.
\end{equation}
Now, the perturbation of the Boltzmann equation yields
\begin{equation}
\label{boltzpert} \imath\omega \delta\!f = -k^2 D \delta\!f + \imath
u k_x \delta\!f -\frac{\imath }{3} \vec{k}\cdot\delta\!\vec{v}
p\frac{\partial f}{\partial p} \;\;\;
\end{equation}
which implies $\delta\! f = 0$ because, for entropy perturbations,
$\omega = uk_x$ and $\vec{k}\cdot\delta\!\vec{v}=0$, as deduced
above. Furthermore, since $\delta\!f =0$, necessarily $\delta\!P_c
=0$, and thus $\delta\!P =0$. It follows that entropy perturbations
and particle perturbations are completely decoupled: in fact,
entropy perturbations are just advected by the zero-th order flow.

In summary, entropy perturbations have the following
characteristics, where we define, for future convenience, $V\equiv
1/\rho$:
\begin{eqnarray}
\label{entropysolution} \delta\!s = \delta\!s_\circ e^{\imath
\omega t-\imath k_y y-\imath k_x x}\nonumber\\ \omega - u k_x = 0
\nonumber \\ \delta\!P = \delta\!P_c = \delta\!f =
\delta\!\vec{v}= 0 \nonumber\\ \frac{\delta\!V}{V} =
\frac{\gamma-1}{\gamma}\frac{m\delta\!s}{k_B}
\end{eqnarray}
where the last equation applies to ideal fluids: $m$ is the
average particle mass and $k_B$ Boltzmann's constant. {}

We turn now to isentropic perturbations, which automatically
satisfy eq. \ref{entropycons}. Mass conservation implies
\begin{equation}\label{mass1}
(\omega-uk_x)\delta\!\rho = \rho \vec{k}\cdot\delta\!\vec{v}
\end{equation}
while momentum conservation implies
\begin{equation}\label{momentum1}
(\omega-uk_x) \delta\!\vec{v} = \frac{\vec{k}}{\rho} (\delta\!P +
\delta\!P_c)\;.
\end{equation}
Let us begin by assuming that the left-hand-side of eq.\ref{mass1}
vanishes; the same is then true for
$\bigtriangledown\cdot\delta\!\vec{v}$. Multiplying eq.
\ref{momentum1} by $\vec{k}\wedge$, we see that
$\bigtriangledown\wedge\delta\!\vec{v} = 0$, unless $\omega = uk_x$.
If the curl vanishes, then so does $\delta\!\vec{v}$, because any
vector with vanishing divergence and curl is a constant, which can
always be set to zero by a suitable choice of reference system. So
we must have $\omega = u k_x$ for perturbations with non-zero
vorticity; we thus find that $\delta\!P+\delta\!P_c = 0$, and then,
again from eq. \ref{boltzpert}, that $\delta\!f = 0$, implying
$\delta\!P = \delta\!P_c = 0$: vorticity perturbations do not couple
to particles either. We have for them:
\begin{eqnarray}\label{vorticitysolution}
\delta\!s = \delta\!\rho = \delta\!P = \delta\!f = \delta\!P_c = 0
\nonumber\\ \omega - uk_x = 0 \nonumber\\
\vec{k}\cdot\delta\!\vec{v} = 0 \;.
\end{eqnarray}

We consider now those perturbations where $\delta\! f$ does not
vanish. We can show that here too there are two distinct classes of
modes: in the first one, $\delta\!f$ is not coupled to the fluid
quantities, while in the second one it is through the term $\vec
k\cdot \delta\!\vec v\neq 0$. The first class of modes, which we
call {\it d-mode}, cannot be obtained directly from eq.
\ref{boltzpert}, for reasons to be made clear shortly. We use
instead the correct form
\begin{equation}\label{nonfourier}
\frac{\partial \delta\! f}{\partial t} = \nabla\cdot(D\nabla
\delta\! f) - u \frac{\partial \delta\!f}{\partial x}\;,
\end{equation}
where we dropped the term $\propto \vec k\cdot \delta\!\vec v$ in
keeping with our desire to find a solution totally uncoupled from
the fluid. The solution of this equation is known from elementary
courses: if $\phi_\circ(x,y,p)$ is the initial distribution function
at time $t=0$ (possibly dependent upon $p$), the solution at later
times for $u = 0$ is
\begin{equation}\label{dmode}
\delta\! f (x,y,t,p) = \int d\!x_0 \int d\!y_0 \;
\phi_\circ(x_0,y_0,p) \frac{1}{4\pi D t}e^{-\frac{(x-x_0)^2}{4D t}}
e^{-\frac{(y-y_0)^2}{4 D t}}\;,
\end{equation}
and the solution for $u\neq 0$ is just $\delta\!f (x-ut,y,p)$. At
the same time, we must make sure that this solution does not ruffle
the fluid: this obviously requires $\delta\! P_c = 0$ at all times.
Now integrating eq. \ref{nonfourier} over $4\pi p^3 v d\!p/3 $ we
find
\begin{equation}
\frac{\partial \delta\! P_c }{\partial t} = \frac{4\pi}{3}
\frac{\partial^2}{\partial x^2} \int D(p) \delta\! f p^3v\; d\!p +
u\frac{\partial \delta\! P_c}{\partial x}\;,
\end{equation}
which clearly shows that, in order to have $\delta\! P_c = 0$
everywhere at all times we need $\delta\! P_c =
 \int D \delta\! f p^3 v\ d\!p = 0$ everywhere at
the initial time. If we now Fourier-expand the initial condition
$\phi_\circ$ with respect to $x,y$, the above conditions become
\begin{equation}\label{justahelp}
\phi_\circ(x,y,p) = \int d\!k_x \int d\!k_y g(p,\vec k) e^{-\imath
k_x x} e^{-\imath k_y y}\;,\;\;\; \int g(p,\vec k) p^3 v \; d\!p =
\int g(p,\vec k) D(p) p^3 v\; d\!p = 0\;.
\end{equation}
This completes the derivation of this purely damped d-mode, which
will not perturb the fluid. Though it may look at this stage like a
mathematical oddity, this mode plays a key role in the matching of
boundary conditions between the upstream and downstream regions. It
is also worth remarking why it cannot be derived from its
Fourier-analyzed counterpart: the solution in question contains a
term $\propto e^{-1/t}$, which does not have a Fourier transform
with respect to $t$.

When $\delta\! P_c \neq 0$, velocity perturbations $\delta\!\vec v
\neq 0$ are induced in the fluid by the non-vanishing particles'
pressure gradient, and eq. \ref{boltzpert}  can be solved for
$\delta\!f$ as
\begin{equation}\label{deltafattheshock}
\delta\!f = -\frac{\delta\!\rho}{\rho} \frac{p}{3}\frac{\partial
f}{\partial p} \frac{\imath(\omega-uk_x)}{\imath(\omega-uk_x) +
k^2 D}\;.
\end{equation}
If we integrate this over $4\pi v p^3 d\!p/3 $, we find
\begin{equation}\label{deltaPc1}
\delta\!P_c = \frac{\delta\!\rho}{\rho}
\int_{p_m}^{p_M}\frac{4\pi}{9} d\!p v p^4 \frac{\partial f}{\partial
p } \frac{-\imath(\omega-uk_x)}{\imath(\omega-uk_x) +k^2 D}\;,
\end{equation}
where we assumed the non-thermal particles to have a minimum ($p_m$)
and a maximum ($p_M$) momentum. It is convenient to introduce a
weighted diffusion coefficient $ \bar D$
defined as follows:
\begin{equation}\label{Dbar}
\frac{1} {1-z \bar D(z)}
\equiv\frac{\int_{p_m}^{p_M}\frac{4\pi}{9}\; d\!p \; v p^4 \;
\frac{\partial f}{\partial p } \;\frac{1}{1 -z D(p)}}
{\int_{p_m}^{p_M}\frac{4\pi}{9} \;d\!p \; v p^4\;\frac{\partial
f}{\partial p}}\;,
\end{equation}
which shows $\bar D$ to be a function of $z$ only.

The above can be simplified a bit by integration by parts. If the
integral were to extend from $0$ to $+\infty$, we would have
\begin{equation}
- \int_0^\infty\frac{4\pi}{9}\; d\!p\; v p^4 \;\frac{\partial
f}{\partial p} = \int_0^\infty\frac{4\pi}{9}\;d\!p\; (\frac{d}{d
p}vp^4)\;f = \frac{4\pi}{3}\int_0^\infty d\!p\; f p^3 v\;(
\frac{4}{3}+\frac{m^2}{3E^2}) \equiv \gamma_c P_c
\end{equation}
where $P_c$ is the particle pressure in the unperturbed flow, and
$\gamma_c$, the effective particle polytropic index, satisfies
$4/3\leq \gamma_c\leq 5/3$. Since however the integral extends only
from $p_m$ to $p_M$, we define
\begin{equation}\label{aux2}{}
- \int_{p_m}^{p_M}\frac{4\pi}{9} \;d\!p v p^4 \;\frac{\partial
f}{\partial p} = \gamma'_c P_c\;.
\end{equation}

Thus eq. \ref{deltaPc1} can now be rewritten as:
\begin{equation}{}\label{deltaPc}
\delta\!P_c = \gamma'_c P_c \frac{\delta\!\rho}{\rho}\frac{1}{1-
\frac{\imath k^2}{\omega-uk_x}\bar D(\frac{\imath
k^2}{\omega-uk_x})}\;.
\end{equation}

We remark that both $\gamma'_c$ and $\bar{D}$ are obtained by
weighing the zero-th order solution, so that they can be
immediately computed as soon as this solution is available.

We can now eliminate $\vec k\cdot \delta\!\vec v$ between eqs.
\ref{mass1} and \ref{momentum1}, and then use $\delta\!P = c_s^2
\delta\!\rho$ (with $c_s$ the sound speed because we are considering
isentropic perturbations) and the equation above to obtain
\begin{equation}
\label{reldisp} \Omega^2 = 1 +\frac{\gamma'_c P_c}{\gamma P}
\frac{1}{1-\frac{\imath k^2}{\omega-uk_x}\bar D(\frac{\imath
k^2}{\omega-uk_x})}
\end{equation}
where we have called
\begin{equation}{}
\Omega \equiv \frac{\omega-uk_x}{k c_s}
\end{equation}{}
the comoving eigenfrequency, in suitably scaled units.

Eq. \ref{reldisp}, together with the definition of $\bar D$ (eq.
\ref{Dbar}), is the sought-after dispersion relation for the
coupled small perturbations in a homogeneous medium.

\subsection{Properties of coupled modes}\label{subsec:coupledmodes}

The case where $D$ is independent of $p$ has been derived before
\citep{ptuskin1981} and coincides with the above equation. Surprisingly,
the existence of this mode was not noticed by \citet{toptygin1999}, even
though a careful treatment of his equations yields precisely the
same dispersion relation as above; this neglect of this mode has
important consequences to be discussed later on.{}

It is best to begin our discussion with the case when $D$, and thus
$\bar D$, is a constant, independent of $p$. The eq. \ref{reldisp}
then reduces to
\begin{equation}\label{reldisp2}{}
(\Omega-\frac{\imath k D}{c_s})(\Omega^2-1) = \Omega \frac{\gamma'_c
P_c}{\gamma P}\;,
\end{equation}
which is a simple polynomial equation of the third order. In this
case two modes reduce to pressure waves, as is most easily seen in
the test-particle regime $P_c = 0$. There is however a third
solution which is only slightly more mysterious: in the test
particle regime these modes represent a local over-density of
particles dissipated by diffusion. When the test particle regime
does not apply, a particle contribution to the sound speed is
introduced by the term $\propto P_c$. This new mode (which we call
the {\it third} mode) is the equivalent of the d-mode when however
the conditions \ref{justahelp} are \emph{not} satisfied: the basic
idea is still the same, the particle overdensity is dissipated, but
since the particle pressure does not vanish, the fluid is
consequently ruffled. Notice also that there are  \emph{two} third
modes, traveling in opposite directions.

The situation is slightly more complex when $D = D(p)$, because one
must solve simultaneously eqs. \ref{reldisp} and \ref{Dbar}. We
begin by considering the limit $k\rightarrow 0$. In this case, and
assuming $\Omega\rightarrow$ a constant, we easily find, to leading
order in $k$, $\Omega^2 = 1+\gamma'_c P_c/(\gamma P)$: as it must,
the dispersion relation allows pure pressure waves, with the
particle pressure providing a correction to the (pure gas) sound
speed, because for large perturbation wavelengths particles are
entrained by the perturbation. We remark that, in this limit,
pressure waves are \emph{supersonic}, in the sense that they are
faster than pressure waves propagating in a pure gas of the same
thermodynamical state, a result already noticed by \citet{toptygin1999}.

\begin{figure}
\plotone{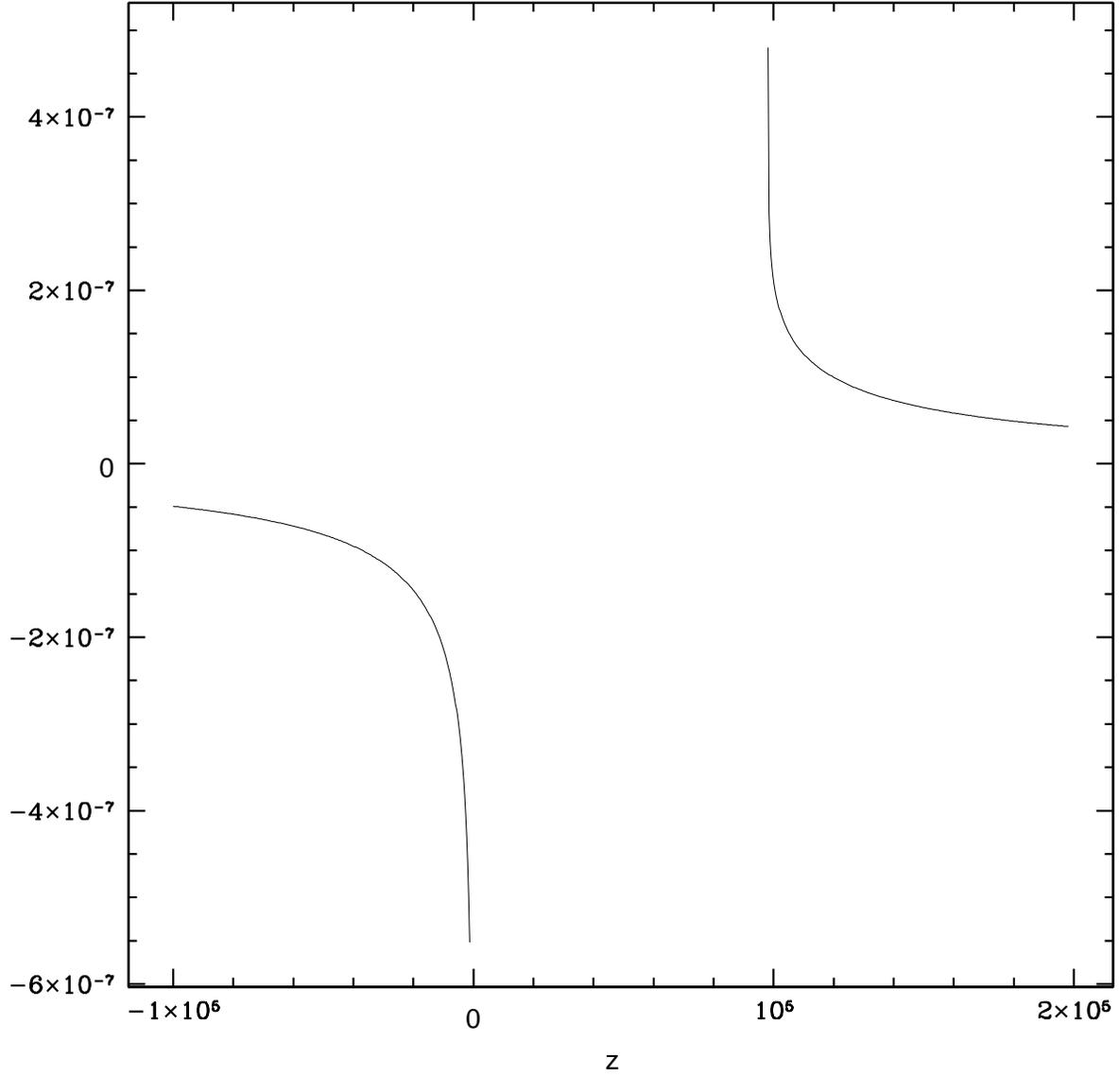}
\caption{The right-hand-side of eq. \ref{approxsol}
as a function of $z$ on the real axis, for one numerical solution
from \citet{amatoblasi2005}
; here,
$p_m = 0.003 mc$, $p_M = 10^5 mc$, $D(p) \propto vp$.}
\label{fig:fz}
\end{figure}

Assuming $\Omega\rightarrow$ a constant, we lost a solution, so we
search for the third mode solution in the form $\Omega = \alpha k+ $
higher order terms in $k$. We find:
\begin{equation}
\Omega = \frac{\imath k \bar D(\imath k^2/(\omega-uk_x))}{c_s}
\frac{1}{1+\frac{\gamma'_c P_c}{\gamma P}}\;.
\end{equation}
Now we use the definition $z \equiv \imath
k^2/(\omega-uk_x) = \imath k/(\Omega c_s)$ to simplify the above to
\begin{equation}
1- z\bar D(z) = -\frac{\gamma'_c P_c}{\gamma P}
\end{equation}
which can be now used with eqs. \ref{Dbar} and \ref{aux2} to obtain
\begin{equation}\label{approxsol}{}
\gamma P = \int_{p_m}^{p_M}\frac{4\pi}{9} \;d\!p v p^4\;
\frac{\partial f}{\partial p }\; \frac{1}{1 -z D(p)}\;,
\end{equation}
where of course $\partial f/\partial p < 0$. As a function of real
$z$, the right-hand side above (where it exists) is easily seen to
be a monotonically decreasing function of $z$, vanishing for
$z\rightarrow\pm\infty$. In any realistic physical problem the
integral must extend from a minimum ($p_m$) to a maximum momentum
$p_M$; since $D(p)$ is expected to be a monotonically increasing
function of $p$, we see that the integral above always exists for $z
< 1/D(p_M) \equiv 1/D_M$, and $z > 1/D(p_m)\equiv 1/D_m$, and it
diverges exactly at $z = 1/D_M$ and $z = 1/D_m$. Thus the integral
on the right-hand side of the equation above spans the whole range
from $0$ to $-\infty$ as $z$ varies between $-\infty$ and
$+1/D(p_m)$, and the range $+\infty$ to $0$ as $z$ varies between
$1/D_m$ and $+\infty$. Somewhere in the range $1/D_m < z < +\infty$
there is the one and only solution of the above equation. An
illustration of the integral on the right-hand side of the previous
equation is shown in Fig. \ref{fig:fz}, for a specific distribution function
from \citet{amatoblasi2005}: the qualitative features of this plot
are generic to all distribution functions we have tried.

\placefigure{fig:fz}

{}

Furthermore, since $z = \imath k^2/(\omega-uk_x)$, and the small
perturbations were assumed to vary as $e^{\imath(\omega-uk_x)t}$,
the result that $z > 0$ implies that all modes are damped by
diffusion, as physical intuition obviously suggests.

The discussion in the opposite limit, $k\rightarrow +\infty$, is
similar. If we assume $\Omega\rightarrow$ a constant, we find the
solution
\begin{equation}
\Omega^2 = 1\;,
\end{equation}
without the correction to the sound speed due to the presence of the
particles' pressure: in the limit $k\rightarrow +\infty$ particles
escape by free streaming, and  do not contribute to the sound speed.

Again, we lost a solution, so we now search for the third mode as
$\Omega = \alpha k +$ lower order terms in $k$. We obtain now:
\begin{equation}
\Omega = \frac{\imath k\bar D(\imath k^2/(\omega-uk_x))}{c_s}
\end{equation}
which can be rewritten as
\begin{equation}
z\bar D(z) = 1\;.
\end{equation}
Comparing this with eq. \ref{Dbar}, we see that the value of $z$ we
are searching for is the one that makes the integral on the
numerator of the right-hand side of eq. \ref{Dbar} diverge.
Following the previous discussion, we see that this is
\begin{equation}\label{aux7}
z = \frac{1}{D(p_m)}\;.
\end{equation}
This is always positive, so that the solution is always damped by
diffusion. This result has a simple physical explanation: when a
small overdensity of particles is generated locally, the time-scale
for damping of this overdensity is dictated by diffusion of the
slowest particles.

\begin{figure}
\plotone{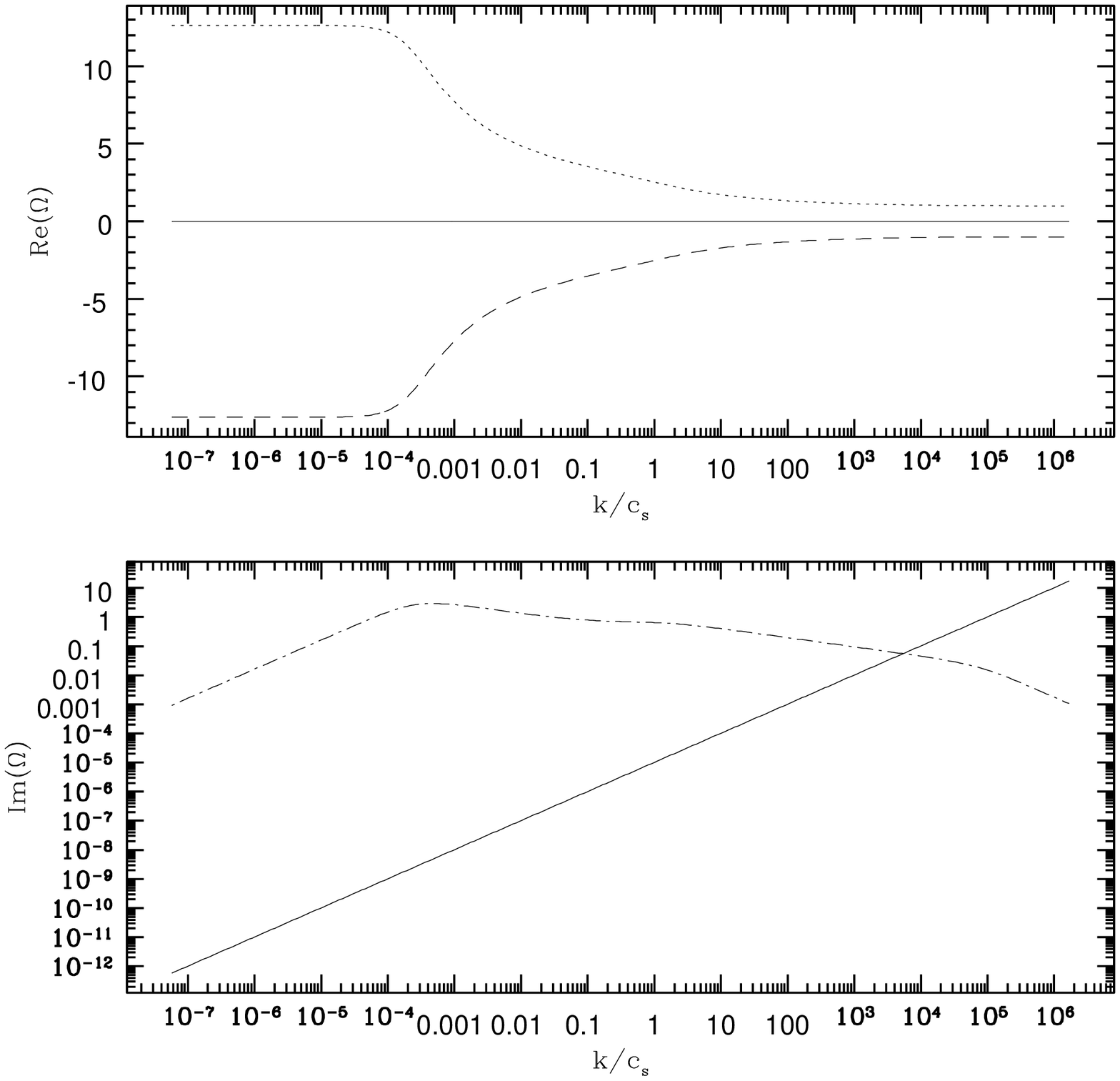}
\caption{Real and imaginary parts for the three
solutions of eq. \ref{reldisp} as a function of perturbation
wave-number $k$, for the same distribution function as in Fig.1 and
$\gamma'_c P_c / (\gamma P)\approx 158.629$ (upstream pressure at
$0^-$). The pressure waves are seen to be damped with the same rate
(and opposite oscillation frequencies), while the third mode is pure
imaginary, corresponding to a pure damping.}
\label{fig:modes}
\end{figure}

Again for illustrative purposes, the real and imaginary parts of
sonic and third modes are displayed in Fig. \ref{fig:modes} for a specific
distribution function from \citet{amatoblasi2005}. Again, the
qualitative features are generic to all models we investigated.

\placefigure{fig:modes}

In summary, we have seen that the introduction of particles modifies
the modes of a homogeneous medium by adding two new modes, one
coupled and one uncoupled to the fluid, both strongly damped by
diffusion.

For future reference, we give the expressions for all small
quantities, in units of $\delta\!P_c$: the link between
$\delta\!P_c$ and $\delta\!\rho$ is obtained from Eqs. \ref{Dbar}
and \ref{aux2}, and the others follow easily. For later convenience,
we use again $V \equiv 1/\rho$ rather than $\rho$:
\begin{eqnarray}\label{boundarycon1}
\label{homogeneoussolution}
\delta\!P_c = \delta\!P_{c\circ}e^{(\imath\omega t - \imath k_y y - \imath k_x x)}\nonumber\\
\frac{\delta\!V}{V} = -\frac{1}{\gamma'_c}\frac{\delta\!P_c}{P_c}
\left(1-\frac{\imath k \bar{D}}{c_s\Omega}\right) = - \frac{\delta\!P_c}{\gamma \,P\,\left( \Omega^2 -1\right)}\equiv
-q\frac{\delta\! P_c}{P_c}\nonumber\\
\delta\!\vec{v} = -\frac{\vec{k}}{k}\frac{c_s}{\Omega}
\frac{\delta\!V}{V} \left(1+\frac{\gamma'_c P_cV}{
c_s^2}\frac{1}{1-\imath\frac{k\bar{D}}{c_s\Omega}}
\right) = -\frac{\vec{k}}{k} \Omega c_s \frac{\delta\!V}{V}\equiv \vec{z}\frac{\delta\! V}{V}\nonumber\\
\frac{\delta\!P}{P} = -\gamma \frac{\delta\!V}{V} \;,
\end{eqnarray}
where the definitions of $\vec z$ and $q$ will come handy later on.
In these equations, $k_x$ must be regarded as a known function of
$\omega$ and $k_y$, specified by eq. \ref{reldisp}.{}

\subsection{Initial value problem}\label{subsec:newnotation}

As a preparation for later work, we discuss the initial value
problem. This has some interest because perturbations in $\delta\!
f$ belonging to the various modes are not mutually orthogonal, and
thus it may appear that initial conditions, especially when given
only in terms of $\delta\! f$, cannot be decomposed into mutually
independent modes. To fix ideas, let us consider a homogeneous
zero-th order solution where all fluid quantities are unperturbed,
but a small perturbation $\delta\! f_0(x,y,p)$ at time $t = 0$ is
given: what are the amplitudes of the four modes that will be
excited (two pressure waves, the third mode, and the d-mode)? First
of course we Fourier-analyze $\delta\! f_0$, calling $a(p)$ its
amplitude. We must then have
\begin{equation}
a(p) = A_{i} \delta\! f_{i}  + g(p)\;,
\end{equation}
where we have introduced some notation that will be useful in the
following: a summation convention over $i$ is understood, the
$A_{i}$'s are the amplitudes of the pressure waves and third modes
for $i=1,2,4$, respectively; $g(p)$ is the amplitude of the d-mode,
as in eq. \ref{justahelp}, and the $\delta\!f_{i}$'s are
\begin{equation}\label{deltafi}
\delta\!f_{i} = -\frac{1}{3}p\frac{\partial f}{\partial
p}\frac{\imath(\omega_{i}-u k_{xi})}{\imath(\omega_{i}-uk_{xi}) +
k_{i}^2 D(p)}\;.
\end{equation}
Comparing this with eq. \ref{deltafattheshock}, it becomes clear
that the mode amplitudes $A_i$'s are simply $\delta\!\rho_i/\rho$.
Here the quantities $\omega_{i}, k_{xi}, k_{i}^2$ are supposed to be
linked by the appropriate branch of the dispersion relation. For
$g(p)$ to represent a proper d-mode, we know that it must satisfy
the two constraints in eq. \ref{justahelp}; thus we derive two
conditions on the mode amplitudes:
\begin{equation}\label{constr1}
\int p^3 v a(p)\; d\! p = A_{i}\int p^3 v \delta\! f_{i}\; d\!p \;
\end{equation}
\begin{equation}\label{constr2}
\int D(p) p^3 v a(p)\; d\! p = A_{i}\int D(p) p^3 v \delta\! f_{i}\;
d\!p \;.
\end{equation}
The last condition can be obtained by satisfying the requirement
that the perturbations to both density and velocity vanish at the
initial time.
With reference to eq. \ref{homogeneoussolution}, we see that the
two conditions, the vanishing of the density and of the velocity,
imply
\begin{equation}
\sum_i A_i  = 0\;,\qquad \Omega_i A_i = 0\;,
\end{equation}
which are two more linear relations which, together with eqs.
\ref{constr1} and \ref{constr2}, determine the $A_i$'s. This simple
example illustrates the importance of the d-mode, a kind of
elephants' graveyard because, once particles join it, the perturbations
they generate can only be dissipated away without ruffling the fluid 
ever again.

\section{Perturbations upstream}\label{sec:pertup}

It is useful to remark that, upstream, entropy  perturbations are
not allowed for arbitrarily dishomogeneous flows, while the same is 
not true for vorticity perturbations.

In fact, the perturbation of the entropy equation is:
\begin{equation}
\frac{\partial\delta\!s}{\partial t} + u(x) \frac{\partial
\delta\!s}{\partial x} = 0
\end{equation}
where we could neglect the term $\delta\! v_x\;d s/d x$ because the
fluid is isentropic in the unperturbed state. From the above, we see
that entropy perturbations are advected by the background flow all
the way from upstream infinity to the shock; we cannot however
accept this, since in our problem all perturbations must have as a
source the shock flapping. Perturbations riding all the way from
upstream infinity belong to perturbations in the boundary
conditions, and are thus irrelevant. Thus the perturbed fluid will
be assumed adiabatic, from now on.

The same argument does not apply to vorticity perturbations when the
fluid is stratified, because they do couple to particle
perturbations: in fact we easily obtain from eq. \ref{momentumcons}
that the equation for the vorticity, $\vec\eta\equiv\nabla\wedge\vec
v$, is:
\begin{equation}
\left(\frac{\partial}{\partial t}+\vec v\cdot\nabla\right)
\frac{\vec\eta}{\rho} =
\left(\frac{\vec\eta}{\rho}\cdot\nabla\right)\vec v
+\frac{1}{\rho^3}\nabla\rho\wedge\nabla\left(P+P_c\right)\;.
\end{equation}{}
In the absence of particles, this equation tells us that vorticity
is exactly ({\it i.e.}, not just to zero or first order)
Lie-advected by the flow because, for adiabatic flows,
$\nabla\rho\wedge\nabla P = 0$. But in the presence of particles
\emph{and} of spatial gradients it is easily seen that the particle
pressure $P_c$ acts as source of vorticity perturbations.{}

In order to make progress, we compute the spatial dependence of
the various modes upstream in the WKBJ approximation: {\it i.e.},
in the limit $k_y\rightarrow\infty$, and in particular $k_y^{-1}\ll
L$, the size of the upstream precursor. In other words, we take all 
perturbed physical quantities to be of the form
\begin{equation}
\delta\!X \approx Q_X(x) e^{\imath\omega t + \imath \int_x^0 W_X(x') d\!x'
-\imath k_y y}\;,
\end{equation}
and all derivatives in the $x$ direction (\ie, perpendicular to the shock)
are considered small when compared with terms proportional to $k_y$. This is 
a perturbation analysis in which 
the transverse wavenumber $k_y$ is assumed large and as a 
consequence the longitudinal wavenumber $\approx W_X$ is also large. 
The presence of a non-constant 
amplitude $Q_X(x)$ is equivalent to keeping the first two terms in
an asymptotic expansion in the small parameter $(k_yL)^{-1}$.
This is often called the physical optics approximation \citep{benderorszag1978}.

This analysis is quite standard, but the amusing thing is that
we don't even need to carry it through. In fact, we shall show
later that the stability analysis requires knowledge of the 
physical quantities immediately before the shock, while knowledge 
of the perturbations run with $x$ further from the shock is
immaterial. We see from the above that all physical quantities 
close to the shock satisfy
\begin{equation}
\delta\!X \approx  Q_X(0) e^{\imath\omega t- \imath W_X(0) x
-\imath k_y x}\;.
\end{equation}
This is precisely the same form that holds in the homogeneous
medium, so that eqs. \ref{homogeneoussolution} and \ref{justahelp}
still hold. Also, the space-time dependence of the d-mode for the half-plane
$x>0$ is derived in Appendix \ref{app:dmode}, assuming spatial 
homogeneity of the background solution; and this, for the argument 
above, applies in the WKBJ limit also to the upstream fluid.
Let us also remark that, in this spatially homogeneous limit, 
vorticity perturbations do not couple to particles: in other 
words, spatial gradients are negligibly small, and thus 
vorticity perturbations, which can only be Lie-advected
from upstream infinity, must vanish identically.

This is the result we need: since we are using a 
WKBJ approximation, we can treat the upstream fluid as if it
were homogeneous, with the values for the physical quantities 
taken to be those immediately before the shock: we call this
the Homogeneous Approximation: it clearly breaks down when the
WKBJ analysis does, which occurs for $k_y^{-1} \approx L$,
the size of the upstream precursor.

\section{Boundary conditions for the particle distribution function}
\label{sec:fitboundary}

Because of diffusion, particles are not restricted to the downstream
part of the flow: there will be a $\delta\!f \neq 0$ also upstream,
so that we need to consider appropriate conditions to match
$\delta\!f$ in the two regions. The first condition is obviously the
continuity across the shock,
\begin{equation}\label{aux19}
\delta\! f_1 = \delta\! f_2\;.
\end{equation}
It is also well-known that the spatial gradient of $\delta\! f$
needs to satisfy a boundary condition at the shock: this is derived
by integrating eq. \ref{boltzmann} on an infinitesimal interval
straddling the shock. The unit vector normal to the surface of the
flapping shock is $\hat{n}=(1,\imath k_y \zeta)$ where $\zeta$ is
the shock corrugation amplitude. So, we obtain:
\begin{equation}\label{integratedbolztmann}
\hat{n}\cdot{\left(D\nabla f\right)}_2 - \hat{n}\cdot{\left(D\nabla f\right)}_1
+ \frac{1}{3}p\frac{\partial f}{\partial p}\;\hat{n}\cdot\left(\vec{v}_2-\vec{v}_1\right)=0\;.
\end{equation}
Writing its first-order linearization and using eq. \ref{aux19} we find:
\begin{eqnarray}\label{firstorder1}
D\left(\frac{\partial \delta\! f}{\partial x}|_2-\frac{\partial \delta\! f}{\partial x}|_1\right)
&+&\imath k_y\zeta D\left(\frac{\partial f}{\partial y}|_2 - \frac{\partial f}{\partial y}|_1\right)\nonumber\\
&+&\frac{1}{3}p\frac{\partial f}{\partial p}\left(\delta\!v_{2x}-\delta\!v_{1x}\right)
+\frac{1}{3}p\frac{\partial \left(\delta\!f\right)}{\partial p}\left(u_2-u_1\right)=0\;,
\end{eqnarray}
We note that ${(\partial f/\partial y)}_2={(\partial f/\partial
y)}_1=0$ because at zero-th order the medium is uniform in
coordinates parallel to the shock surface. The result is:{}
\begin{equation}\label{aux20}
D\left(\frac{\partial \delta\! f}{\partial x}|_2-\frac{\partial
\delta\! f}{\partial x}|_1\right) +\frac{1}{3}p\frac{\partial
f}{\partial p}\left(\delta\!v_{2x}-\delta\!v_{1x}\right)
+\frac{1}{3}p\frac{\partial \left(\delta\!f\right)}{\partial
p}\left(u_2-u_1\right) = 0
\end{equation}
Eqs. \ref{aux19} and \ref{aux20} are the appropriate boundary
conditions for our problem.

We now show how to satisfy the boundary conditions, eqs. \ref{aux19}
and \ref{aux20} at the shock. Using the notation introduced in Sect.
\ref{subsec:newnotation}, we know that $\delta\! f_+$ on the downstream
side of the shock (the factor $e^{\imath\omega t-\imath k_y y}$ will
be omitted for simplicity in this subsection) satisfies
\begin{equation}\label{decomp1}
\delta\! f_+ = A_{id}\delta\!f_{id}+ g_d\;.
\end{equation}
Please notice that both downstream and upstream the summation is
over 2 modes (a pressure wave and a third mode). In fact, we know on 
the one hand that particles perturbations are not coupled to 
entropy and vorticity perturbations, while, on the other hand,
we know that pressure and third modes, for given values of $k_y$ and
$\omega$, exist for opposite values of $k_x$ (see eq. \ref{reldisp}),
and thus at least one pressure mode and one third mode will diverge
exponentially toward infinity: this is unacceptable because we
are studying flow instabilities, not variations in the boundary
conditions at infinity. This discussion will be completed in Section 
\ref{sec:eigenf}.

On the upstream side we have analogously
\begin{equation}\label{decomp2}
\delta\! f_- = A_{iu}\delta\!f_{iu}+ g_u\;,
\end{equation}
The continuity of $\delta\! f$ at the shock allows us to derive a
relationship between the $g$'s:
\begin{equation}\label{gd}
g_d = g_u + A_{iu}\delta\! f_{iu}- A_{id}\delta\! f_{id}\;.
\end{equation}

Using
\begin{eqnarray}
\frac{\partial \delta\!f_+}{\partial x}
&=&-\imath A_{id}k_{xid}\delta\!f_{id}+k_{xd}g_d
=-\imath A_{id}k_{xid}\delta\!f_{id}+k_{xd}g_u+A_{iu}k_{xd}\delta\!f_{iu}-A_{id}k_{xd}\delta\!f_{id}\\
\frac{\partial \delta\!f_-}{\partial x}
&=&-\imath A_{iu}k_{xiu}\delta\!f_{iu}+k_{xu}g_u
\end{eqnarray}
into eq. \ref{aux20} we find
\begin{eqnarray}
D (k_{xd}-k_{xu})g_u
+\frac{u_2-u_1}{3}\frac{\partial g_u}{\partial\ln p}
=
&-&\frac{u_2-u_1}{3}A_{iu}\frac{\partial\delta\! f_{iu}}{\partial\ln p}
-\frac{1}{3}\frac{\partial f}{\partial\ln p}(\delta\! v_{2x}-\delta v_{1x})\nonumber\\
&-&D A_{iu}\delta\!f_{iu}(\imath k_{xiu}+k_{xd})
+D A_{id}\delta\!f_{id}(\imath k_{xid}+k_{xd})
\end{eqnarray}
which we regard as an equation for $g_u(p)$, whose solution can be
written as
\begin{equation}
g_u(p) = C w_C(p) + A_{iu} w_{iu}(p) + A_{id}
w_{id}(p)+(\delta\!v_{2x}-\delta\!v_{1x})w_0(p)\;,
\end{equation}
where the functions $w$'s are derived in the Appendix \ref{app:funcw}.

We can now impose the conditions \ref{justahelp} on $g_u$,
obtaining:
\begin{equation}\label{link1}
C \int p^3 v w_C d\!p + A_{iu}\int p^3 v w_{iu} d\!p + A_{id} \int
p^3 v w_{id} d\!p +(\delta\!v_{2x}-\delta\!v_{1x})\int p^3 v w_0(p)
d\!p= 0\;,
\end{equation}
\begin{eqnarray}\label{link2}
C \int p^3 v w_C D d\!p + A_{iu}\int p^3 v w_{iu} D d\!p &+& A_{id}
\int p^3 v w_{id} D d\!p \nonumber\\
&+&(\delta\!v_{2x}-\delta\!v_{1x})\int p^3 v D w_0(p) d\!p= 0\;,
\end{eqnarray}
And now we can impose the very same conditions on $g_d$, eq.
\ref{gd}, to obtain:
\begin{equation}\label{link3}
A_{id}\int p^3 v \delta\! f_{id} d\!p - A_{iu} \int p^3 v
\delta\!f_{iu} d\!p = 0
\end{equation}
\begin{equation}\label{link4}
A_{id}\int p^3 v \delta\! f_{id} D d\!p - A_{iu} \int p^3 v
\delta\!f_{iu} D d\!p = 0
\end{equation}
This set of four linear equations in five unknowns ($C$ and the
$A_i$'s) can be solved in terms of one of them, say $A_{1d}$, the
pressure wave downstream.

We thus see that the conditions at the shock, plus knowledge of the
modes, allows us to determine the amplitude of all modes (except the
vorticity and entropy modes downstream) in terms of the amplitude of
the pressure wave downstream. Remembering eq.
\ref{homogeneoussolution}, we now see that all fluid quantities at
the shock upstream are determined in terms of the amplitude of this
very same wave; as for the two d-modes, their explicit space- and
time-dependence is not needed, but is given for completeness' sake
in Appendix \ref{app:dmode}. Now, before discussing how to fix $\omega$, the shock
eigenfrequency, we briefly summarize how to perturb the standard
Rankine--Hugoniot conditions.

\section{The fluid conditions at the shock}\label{sec:fluifcond}

We follow closely the discussion in \citet[Section 90]{landaulifshitz1987}, which requires only a small adaptation to our problem.
In their consideration of the corrugational instability, in fact,
there were no perturbations on the upstream side of the shock:
deviations from the unperturbed state were generated by the
corrugation of the shock surface (and its motion), and led to
non-zero perturbations only downstream. In our problem, instead,
the particles crossing the shock manage to generate new
perturbations on the upstream side, leading to a slight modification
to the Rankine--Hugoniot conditions. From now on, we indicate with
the subscript 1 the quantities on the upstream side of the shock,
and with the subscript 2 those on the downstream side. Also, to make
contact with \citeauthor{landaulifshitz1987}' work easier, we use as a variable
$V \equiv1/\rho$ rather than the density $\rho$ directly.

We consider a small-amplitude corrugation of the shock surface,
away from the $x=0$ plane, by a small displacement of the form:
\begin{equation}\label{corrugation}
\zeta = \zeta_\circ e^{\imath\omega t - \imath k_y y}
\end{equation}
with respect to which the unit vectors parallel $\hat{t}$ and
normal $\hat{n}$ to the surface have components in the $xy$ plane:
\begin{equation}\label{vectors}
\hat{t} = (-\imath k_y \zeta,1)\;\;;\;\; \hat{n} = (1,\imath
k_y\zeta)
\end{equation}
while the surface speed in the direction normal to the surface,
with the respect to the reference frame of the unperturbed shock,
is:
\begin{equation}\label{surfacespeed}
\vec{q}\cdot\hat{n} = \imath\omega\zeta\;.
\end{equation}
All quantities are, of course, accurate to first order only.

The first two Rankine--Hugoniot conditions to be perturbed involve
the fluid speed, and they are the continuity of the fluid speed
parallel to the shock surface, and the discontinuity of the
perpendicular component in terms of perturbed pressure and density
\citep[eq. 85.7 of][]{landaulifshitz1987}. We have:
\begin{eqnarray}\label{firsttwo}
(\vec{v}_1 +\delta\!\vec{v}_1)\cdot \hat{t} = (\vec{v}_2
+\delta\!\vec{v}_2)\cdot \hat{t}\nonumber\\
(\vec{v}_1+\delta\!\vec{v}_1)\cdot\hat{n} -
(\vec{v}_2+\delta\!\vec{v}_2)\cdot\hat{n} = \sqrt{(P_2+\delta\!P_2
- P_1-\delta\!P_1)(V_1+\delta\!V_1-V_2-\delta\!V_2)}
\end{eqnarray}
whose first-order linearizations are:
\begin{eqnarray}\label{firsttwosolutions}
\delta\!v_{2y}-\delta\!v_{1y} = \imath k_y \zeta(v_2-v_1)
\nonumber\\ \delta\!v_{2x}-\delta\!v_{1x} = \frac{1}{2} (v_2-v_1)
\left( \frac{\delta\!P_2-\delta\!P_1}{P_2-P_1}-
\frac{\delta\!V_2-\delta\!V_1}{V_1-V_2}\right)\;.
\end{eqnarray}

The next equation to be perturbed is the shock adiabat, which is
convenient because it is independent of all speeds involved. For a
polytropic gas like the one we are considering, the unperturbed
Hugoniot adiabat is given by \citep[eq. 89.1]{landaulifshitz1987}:
\begin{equation}\label{hugoniot}
\frac{V_2}{V_1} =
\frac{(\gamma+1)P_1+(\gamma-1)P_2}{(\gamma-1)P_1+(\gamma+1)P_2}
\end{equation}
whose perturbation yields
\begin{equation}
\label{partialsolution} \frac{\delta\!V_2}{V_2} =
\frac{\delta\!V_1}{V_1}+h\left(
\frac{\delta\!P_1}{P_1}-\frac{\delta\!P_2}{P_2} \right)
\end{equation}
where we have defined
\begin{equation}
h \equiv
\frac{4\gamma}{((\gamma+1)+(\gamma-1)P_2/P_1)((\gamma-1)P_1/P_2+(\gamma+1))}
\end{equation}
while the ratio $P_2/P_1$ can be expressed as
\begin{equation}
\frac{P_2}{P_1} = \frac{2\gamma M_1^2-(\gamma-1)}{\gamma+1}
\end{equation}
with $M_1$ the Mach number of the upstream fluid.

We need one more equation: we can use  the equation relating the
mass flux to the discontinuities in pressure and density: eq. 85.6
of \citet{landaulifshitz1987} recites:
\begin{equation}
j^2 = \frac{P_2-P_1}{V_1-V_2}
\end{equation}
where $j$ is the mass flux. When the shock surface is perturbed,
the above becomes
\begin{equation}
\label{fakemassflux}
\frac{((\vec{v}_1+\delta\!\vec{v}_1)\cdot\hat{n}-\vec{q}\cdot\hat{n})^2}
{(V_1+\delta\!V_1)^2} =
\frac{P_2+\delta\!P_2-P_1-\delta\!P_1}{V_1+\delta\!V_1-V_2 -
\delta\!V_2}\;.
\end{equation}
The perturbation to first order of the above yields
\begin{equation}\label{massflux}
\frac{2\delta\!v_{1x}}{v_1}-\frac{2\imath\omega\zeta}{v_1}
-\frac{2\delta\!V_1}{V_1} =
\frac{\delta\!P_2-\delta\!P_1}{P_2-P_1}-\frac{\delta\!V_1-\delta\!V_2}{V_1-V_2}
\end{equation}

Lastly, the condition that $\delta\!f$ be continuous across the
shock  has already been imposed, see eq. \ref{aux19}.

\section{The equation for the perturbation eigenfrequency}\label{sec:eigenf}

We discuss here first how this problem differs from the one
without particles, keeping in mind that 
we are interested in the stability of the shock flapping,
so that all modes present must have this flapping as their source:
we cannot allow modes to come in from spatial infinity, because
this amounts to perturbation of the boundary conditions, not 
of the shock geometry.

When we can neglect the particles'
pressure, we know that there can be no perturbations upstream, since
they are either generated at upstream infinity (in which case we
would not be treating the case of shock instability) or, if
generated at the shock, they cannot propagate away from it fast
enough (the shock is supersonic). In the presence of particles the
situation changes because particles can diffuse back to the upstream
region, so that a perturbation $\delta\! f$ generated downstream can
return to upstream and perturb the fluid quantities. The situation
is even more remarkable when one notices that, in this way, there
can be a generation of pressure waves (even though they are just
sonic in a supersonic medium) in the upstream region. The reason is
shown in eq. \ref{momentumcons}: the gradient in particles' pressure
is a source of perturbations, and since particles scatter a \emph{finite} ({\it i.e.}, non-zero) distance from the shock, there is no
obvious reason why even sonic perturbations should not be generated.
The impossibility of having sonic perturbations in a supersonic
medium arises only when the point of generation of the perturbations
is the shock itself, not a finite distance from it.

We must also remark that the presence of particles shuffling between
downstream and upstream, and {\it viceversa}, has important
consequences also for the kind of waves present downstream. In fact,
while in the absence of particles the only waves present can be
those shed by the shock, {\it i.e.} entropy, vorticity and pressure
disturbances propagating to downstream infinity, when particles are
included in the picture we find, by complete analogy with the
argument above for the upstream region, that they can seed 
the third mode and the d-mode.

We have seen in Section \ref{sec:fitboundary} that the amplitude of all
modes (and thus of all physical quantities) upstream, and of
the third mode downstream, can be expressed in terms of a single 
free parameter, the amplitude of pressure waves downstream. Thus
all amplitudes are fixed, except for three modes downstream (entropy,
vorticity, pressure) and the shock displacement $\zeta$; but we also
have the four perturbed Rankine--Hugoniot conditions, which can be
expressed in terms of these four quantities. The vanishing of 
determinant, once substitution of amplitudes for the upstream
modes, and for the third mode downstream by means of the relations
in Section \ref{sec:fitboundary} has been made, thus fixes the eigenfrequency.

A word of caution is in order: we have tacitly assumed that the pressure
wave propagating away from the shock in the downstream region is
also the wave with the physically acceptable ({\it i.e.}, decreasing)
behaviour at downstream infinity. This is not certain: in fact, even
in the standard case with no particles, a proper mode exists when
the mode explodes exponentially in time  \emph{and} the pressure wave
propagating away from the shock dies down at downstream infinity \citep{landaulifshitz1987}. This condition can only be checked {\it a posteriori},
{\it i.e.} after having found numerically the value of $\omega$.

We now explicitly derive the set of equations whose determinant must
vanish, which is what fixes $\omega$ for a given $k_y$. 

In the four Rankine--Hugoniot equations there appear the total perturbations of specific volume, pressure and velocity. They can be written as sum of the respective perturbation for each mode. Downstream we have:
\begin{eqnarray}
\delta\!V_2&=&\delta\!V_{2s}+\delta\!V_{2p}+\delta\!V_{2t}\;,\\
\delta\!\vec{v}_2&=&\delta\!\vec{v}_{2v}+\delta\!\vec{v}_{2p}+\delta\!\vec{v}_{2t}\;,\\
\delta\!P_2&=&\delta\!P_{2p}+\delta\!P_{2t}\;,
\end{eqnarray}
where the subscripts $s$, $v$, $p$ and $t$ label quantities for entropy, vorticity, pressure waves and third mode.
Upstream there are neither entropy nor vorticity perturbations:
\begin{eqnarray}
\delta\!V_1&=&\delta\!V_{1p}+\delta\!V_{1t}\;,\\
\delta\!\vec{v}_1&=&\delta\!\vec{v}_{1p}+\delta\!\vec{v}_{1t}\;,\\
\delta\!P_1&=&\delta\!P_{1p}+\delta\!P_{1t}\;.
\end{eqnarray}
We must use these expressions in the perturbed Rankine--Hugoniot. Furthermore, we can write one of the two components of $\delta\!\vec{v}_{2v}$ in terms of the other one, through $\vec{k}_{2v}\cdot\delta\!\vec{v}_{2v}=0$, where $k_{2vy}\equiv k_y$ and $k_{2vx}=\omega/v_2$ (see eqs. \ref{vorticitysolution}). Then we can simplify our system by eliminating the shock displacement $\zeta$, the remaining component of the velocity perturbation due to vorticity, and $\delta\!V_{1s}$. One equation remains, linking only pressure waves and third modes' perturbations:
\begin{eqnarray}
 \omega \,{P_2}\,\left( {v_1} - {v_2} \right) \,{V_1}\,\left\{  {V_1}{P_1}\,\left( {\omega }^2 - {{k_y}}^2\,{v_1}\,{v_2} \right) 
\right. \nonumber\\ \left.
 - 
       \left[ {\omega }^2\,\left( \left( 1 + h \right) \,{P_1} - h\,{P_2} \right)  + {{k_y}}^2\,\left( \left(  h -1\right) \,{P_1}  - h\,{P_2} \right) \,{v_1}\,{v_2} \right] \,{V_2} \right\}
      \,(\delta\!P_{1p} +\delta\!P_{1t})\nonumber\\
-
\omega \,{P_1}\,\left( {v_1} - {v_2} \right) \,{V_1}\,
   \left\{ {V_1}{P_2}\,\left( {\omega }^2 - {{k_y}}^2\,{v_1}\,{v_2} \right)  
\right. \nonumber\\ \left.
 - 
     \left[ {\omega }^2\,\left( h\,{P_1} - \left(  h -1\right) \,{P_2} \right)  + {{k_y}}^2\,\left( h\,{P_1} - \left( 1 + h \right) \,{P_2} \right) \,{v_1}\,{v_2} \right] \,{V_2} \right\} \,
   (\delta\!P_{2p} +\delta\!P_{2t})\nonumber\\
+
\omega \,{P_1}\,\left( {P_1} - {P_2} \right) \,{P_2}\,\left( {v_1} - {v_2} \right) \,\left( {\omega }^2 - {{k_y}}^2\,{v_1}\,{v_2} \right) \,
   \left( {V_1} - {V_2} \right) \,(\delta\!V_{1p}+\delta\!V_{1t}) \nonumber\\
- 
  2\,\omega \,{P_1}\,\left( {P_1} - {P_2} \right) \,{P_2}\,\left( {\omega }^2 + {{k_y}}^2\,{v_2}\,\left( -{v_1} + {v_2} \right)  \right) \,{V_1}\,\left( {V_1} - {V_2} \right) \,
   (\delta\!v_{1px}+\delta\!v_{1tx})\nonumber\\
- 2\,{\omega }^2\,{k_y}\,{P_1}\,\left( {P_1} - {P_2} \right) \,{P_2}\,{v_2}\,{V_1}\,\left( {V_1} - {V_2} \right) \,(\delta\!v_{1py}+\delta\!v_{1ty})\nonumber\\
 + 2\,{\omega }^3\,{P_1}\,\left( {P_1} - {P_2} \right) \,{P_2}\,{V_1}\,\left( {V_1} - {V_2} \right) \,   (\delta\!v_{2px}+\delta\!v_{2tx}) \nonumber\\
+ 2\,{\omega }^2\,{k_y}\,{P_1}\,\left( {P_1} - {P_2} \right) \,{P_2}\,{v_2}\,{V_1}\,\left( {V_1} - {V_2} \right) \,  (\delta\!v_{2py}+  \delta\!v_{2ty})=0\;.
\end{eqnarray}
Now, eqs. \ref{homogeneoussolution} hold both for pressure waves and third modes. We can use them to write the above equation in terms of volume perturbations. We remark that each mode has its own $\vec{z}$ and its own $k_x$ but, for given $k_y$ and $\omega$, they are fixed by the dispersion relation, eq. \ref{reldisp}. We obtain:

\begin{eqnarray}\label{eigenfrequency}
\sum_{i=p,t}{\frac{\delta\!V_{1i}}{V_1}\,\left[  \left( \left( \gamma -1 \right) \,{P_1} + {P_2} \right) \,
          \left( {v_1} - {v_2} \right) \,\left( \omega^2 - {{k_y}}^2\,{v_1}\,{v_2} \right) \,\left( {V_1} - {V_2} \right)  
\right.} \nonumber\\ \left.
  - 
       h\,\gamma \,\left( {P_1} - {P_2} \right) \,\left( {v_1} - {v_2} \right) \,\left( {\omega }^2 + {{k_y}}^2\,{v_1}\,{v_2} \right) \,{V_2} 
\right. \nonumber\\ \left.
 + 
       2\,\left( {P_1} - {P_2} \right) \,\left( {V_1} - {V_2} \right) \,\left(  \left( {\omega }^2 + {{k_y}}^2\,{v_2}\,\left(  {v_2}-{v_1}  \right)  \right) \,
             z_{1ix}  + \omega \,{k_y}\,{v_2}\,z_{1iy} \right)  \right] \nonumber\\
\nonumber\\ 
 -
\sum_{i=p,t}{\frac{\delta\!V_{2i}}{V_2}\,\left[ \gamma\,{P_2}\,\left( {v_1} - {v_2} \right) \,\left( {\omega }^2 - {{k_y}}^2\,{v_1}\,{v_2} \right) \,
        \left( {V_1} - {V_2} \right) 
\right.} \nonumber\\ \left.
 -
 h\,\gamma \,\left( {P_1} - {P_2} \right) \,\left( {v_1} - {v_2} \right) \,\left( {\omega }^2 + {{k_y}}^2\,{v_1}\,{v_2} \right) \,{V_2}
\right. \nonumber\\ \left.
 + 
       2\,\omega \,\left( {P_1} - {P_2} \right) \,\left( {V_1} - {V_2} \right) \,\left( \omega \,z_{2ix} + {k_y}\,{v_2}\,z_{2iy} \right)  \right]
=0\;,
\end{eqnarray}
where the sum is over the pressure mode and the third mode. Now, recalling the definition of $A_i$, we see that $A_i=-\delta\!V_i/V$ (see also the discussion following eq. \ref{deltafi}). Then we have the four equations \ref{link1}, \ref{link2}, \ref{link3}, \ref{link4}, plus eq. \ref{eigenfrequency}, for a total of five linear homogeneous equations in five unknowns: four $A_i$ and $C$. The system has non-trivial solution only if its determinant vanishes. This condition determines the eigenfrequency of the system.

\section{Results}\label{sec:results}

\begin{figure}
\plotone{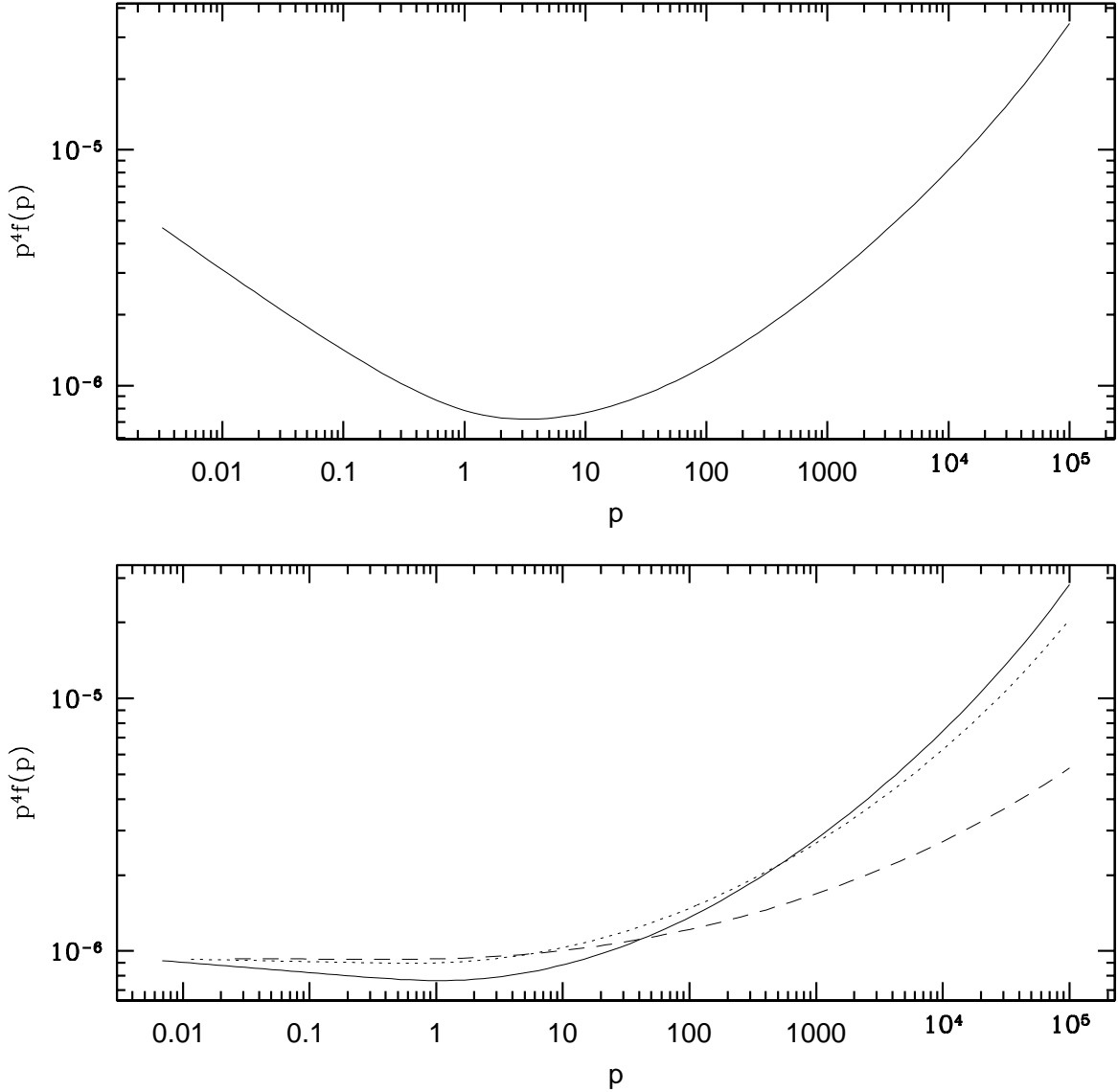}
\caption{Particle distribution function $f$ as a function of momentum $p$, measured in units of $mc$. The value of $f$ is divided by the fluid density $V_0$ at upstream infinity and multiplied by $p^4$. Upper panel: single solution, obtained by \citet{amatoblasi2005} for a shock with Mach number at upstream infinity $M_1=100$. Results of our stability analysis are reported in Fig. \ref{fig:onesolution}. Lower  panel: triple solution, obtained by \citet{amatoal2008} for a shock with Mach number at upstream infinity $M_1=100$. Results of our stability analysis for particle distribution in solid, dotted and dashed lines are reported, respectively, in Figs. \ref{fig:asolution}, \ref{fig:bsolution} and \ref{fig:csolution}.}
\label{fig:fdistr}
\end{figure}

In this section we apply our stability analysis to two shock solutions: a single solution (Fig. \ref{fig:fdistr}, upper panel) by \citet{amatoblasi2005} and a multiple solution (Fig. \ref{fig:fdistr}, lower panel) by \citet{amatoal2008}.

As we stated above, instability takes place when perturbations exist and grow exponentially with time. Furthermore,
they must decay away from the shock. These conditions may be summarized as follows:
\begin{equation}\label{eq:instcond}
\Im{(\omega)}<0\,,\qquad\Im{(k_{xid})}<0\,,\qquad\Im{(k_{xiu})}>0\,,
\end{equation}
where the subscript $i$ indicates that these conditions must hold for all waves. 

For illustrative purposes, first we consider the solution of D'yakov's equation \citep[see][eq. 90.10]{landaulifshitz1987}:
\begin{equation}\label{eq:eigenfrequencyl}
\frac{2\omega u_2}{u_1}\left(k_y^2+\frac{\omega^2}{u_2^2}\right)-\left(\frac{\omega^2}{u_1 u_2}+k_y^2\right)\left(\omega-u_2 k_x\right)(1+h_L)=0\,,
\end{equation}
which fixes the shock eigenfrequency in the linear regime, with
\begin{equation}
h_L\equiv j^2{\left.\frac{\partial V_2}{\partial P_2}\right|}_{V_1,P_1}\,.
\end{equation}

We apply D'yakov's analysis to a test particle solution for a strong shock (upstream Mach number $M_1\rightarrow\infty$) in a polytropic fluid with index $\gamma=5/3$. From eqs. 89.6 and 89.9 in \citet{landaulifshitz1987} we obtain, respectively, the compression factor $R=V_1/V_2=4$ and the downstream Mach number $1/\sqrt{5}$. The downstream sound speed is $c_{s2}=u_2/M_2=u_1/RM_2=u_1\sqrt{5}/4$. In such system sound waves can propagate away from the shock only downstream. The dispersion relation for such perturbations is straightforward:
\begin{equation}\label{eq:displan}
{\left(\omega-\frac 1 4 u_1 k_x\right)}^2=\frac 5 {16} u_1^2 \left(k_y^2+k_x^2\right)\,,
\end{equation}
where $k_x$ is the $x$-component of the sound wave. To write down the eigenfrequency equation for corrugations
of the shock we need to calculate $h_L$. This task is straightforward for a strong shock because the upstream pressure vanishes. As a consequence, the derivative ${\left(\partial V_2/\partial P_2\right)}_{V_1,P_1}$ vanishes too (see eq. \ref{hugoniot}) and $h_L=0$. Eq. \ref{eq:eigenfrequencyl} becomes:
\begin{equation}\label{eq:eiglan}
4\omega^3 - \frac 1 2 \omega u_1^2 k_y^2 + u_1 k_x\left( {\omega }^2 + \frac 1 4 u_1^2 k_y^2 \right)  = 0\,.
\end{equation}
Eqs. \ref{eq:displan} and \ref{eq:eiglan} form a system to be solved with respect to $\omega$ and $k_x$ for a given value of $k_y$. Since these equations are third-degree homogeneous in $\omega$, $k_y$, $k_x$, if $(\omega,k_x)$ is a solution for a given $k_y$, then $(\lambda\omega,\lambda k_x)$ is a solution for $\lambda k_y$. Thus the problem is completely solved once all the solutions for a given $k_y$ are found. Let be $k_y=1$. Also the upstream fluid speed $u_1$ can be set to 1 by redefining the ratio between the units of measure of frequency and wave number. We calculated all the solutions of the above equations, for these values of $u_1$ and $k_y$, and display them in Tab. \ref{tab:sollan}.

\placetable{tab:sollan}

\begin{table}
\centering
\begin{tabular}{*{2}{c}}
\hline
$\omega$		&	$k_x$			\\
\hline	
$\frac 1 2$		&	$-\frac 1 2$		\\
$-\frac 1 2$		&	$\frac 1 2$		\\
$\frac{\sqrt{5}}2$	&	$-\frac{3\sqrt{5}}2$	\\
$-\frac{\sqrt{5}}2$	&	$\frac{3\sqrt{5}}2$	\\
$\frac \imath 4$	&	$\imath$		\\
$-\frac \imath 4$	&	$-\imath$		\\
\hline
\end{tabular}
\caption{Solutions of eqs. \ref{eq:displan} and \ref{eq:eiglan} for $k_y=u_1=1$.}
\label{tab:sollan}
\end{table}

The first four solutions must be discarded because they correspond to waves propagating from downstream infinity to the shock. The fifth solution must be discarded too because it diverges exponentially at downstream infinity. More intriguing is the last solution. It seems to satisfy all the requirements in order to be a real physical solution and it actually satisfies conditions \ref{eq:instcond}. Is it a real instability? No, because it represents a pressure perturbation advected by the fluid since $\omega-u k_x = 0$. From eq. 90.5 in \citet{landaulifshitz1987} we see that such a perturbation should have $\delta\!P=0$. This is clearly absurd: we assumed $k_x$ to be the wave vector of a pressure perturbation in order to obtain eq. \ref{eq:eigenfrequencyl} but we have now found a solution with vanishing pressure perturbation. Note that the fifth solution is affected by this problem too.

Summarizing, we solved the equation for the shock eigenfrequency in the test particle regime. We found six solutions, but 
none has physical meaning, since they correspond to sound waves either propagating from downstream infinity to the shock surface or with vanishing pressure perturbation.

Below we apply our theory to two particular shock structures (one with multiple solution) in order to search for an instability. We do not want to carry out a systematic analysis of a set of solutions but we shall just show how our \emph{machinery} works. For the sake of completeness in Fig. \ref{fig:fdistr} we report the particle distributions $f$ we used in our calculations. 

\placefigure{fig:fdistr}

We adopted a system of units of measure so that the following three quantities equal 1: the speed of light, the proton mass and the numerical constant of the Bohm diffusion coefficient, \ie, $D_p=v(p)p$.  Below everything will be expressed in these units.

\begin{figure}
\plotone{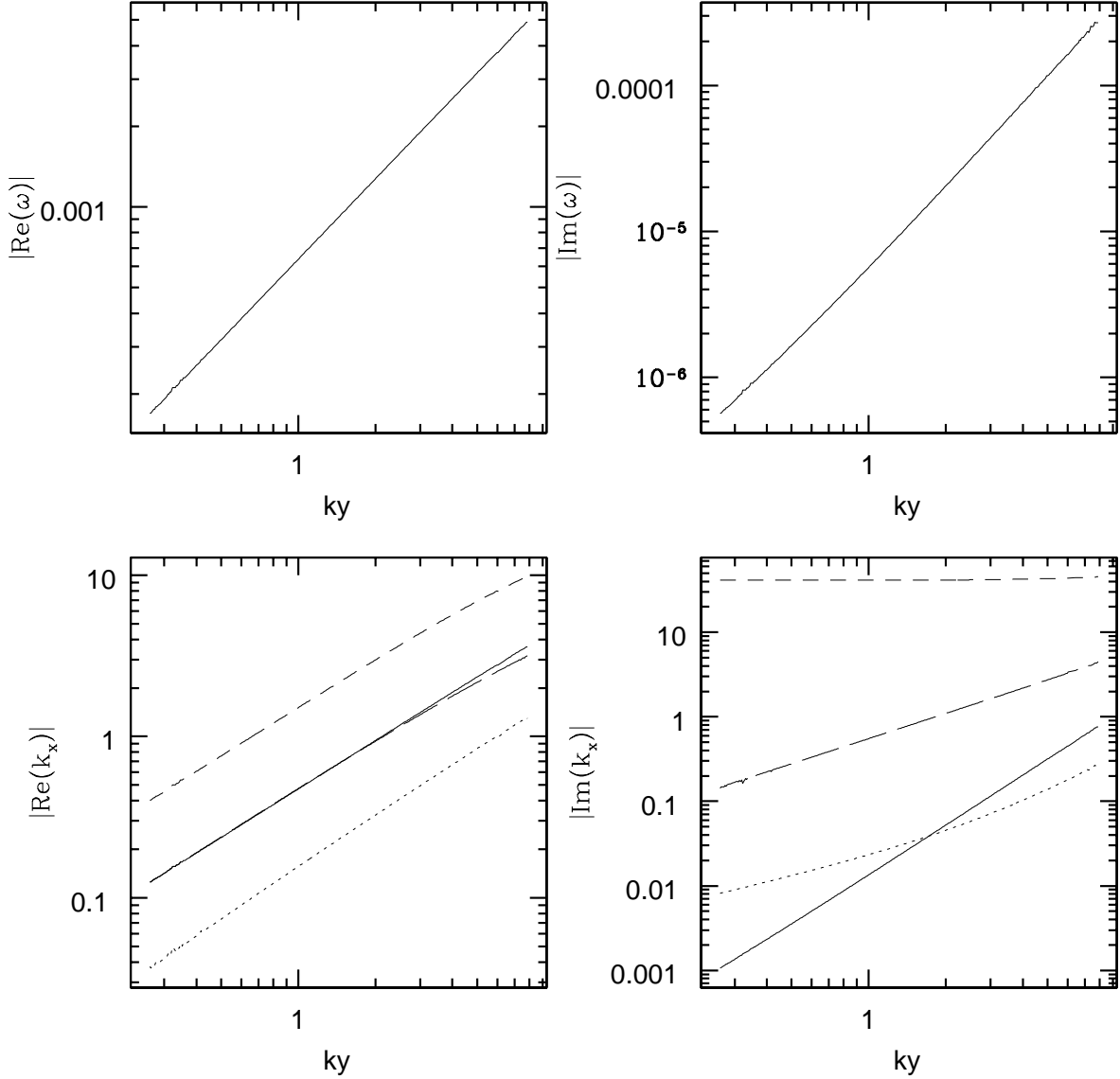}
\caption{Solution of our equations as a function of $k_y$, for the shock structure with accelerated particles' distribution plotted in Fig. \ref{fig:fdistr}, upper panel. Upper panels: absolute values of real part (left panel) and imaginary part (right panel) of the eigenfrequency $\omega$. Lower panels: absolute values real parts (left panel) and imaginary parts (right panel) of the $x$-component of the wave vectors $k_{xpu}$ (pressure mode upstream, dotted lines), $k_{xtu}$ (third mode upstream, solid lines), $k_{xpd}$ (pressure mode downstream, long-dashed lines), $k_{xtd}$ (third mode downstream, short-dashed lines). Signs of these quantity are reported in Tab. \ref{tab:signs}.}
\label{fig:onesolution}
\end{figure}
\begin{figure}
\plotone{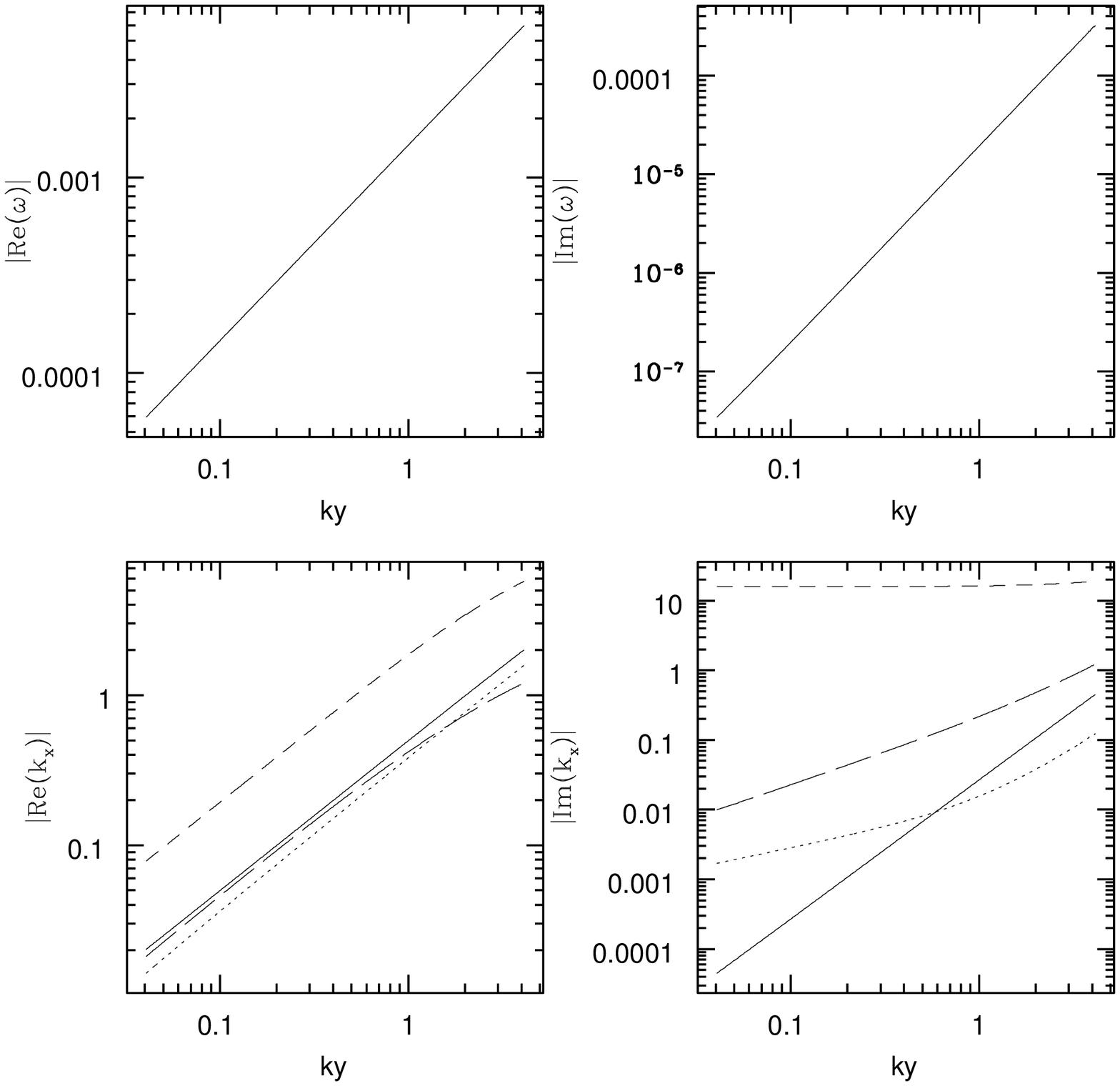}
\caption{Solution of our equations as a function of $k_y$, for the shock structure with accelerated particles' distribution plotted in Fig. \ref{fig:fdistr}, lower panel, solid line. Please refer to caption of Fig. \ref{fig:onesolution} for details.}
\label{fig:asolution}
\end{figure}
\begin{figure}
\plotone{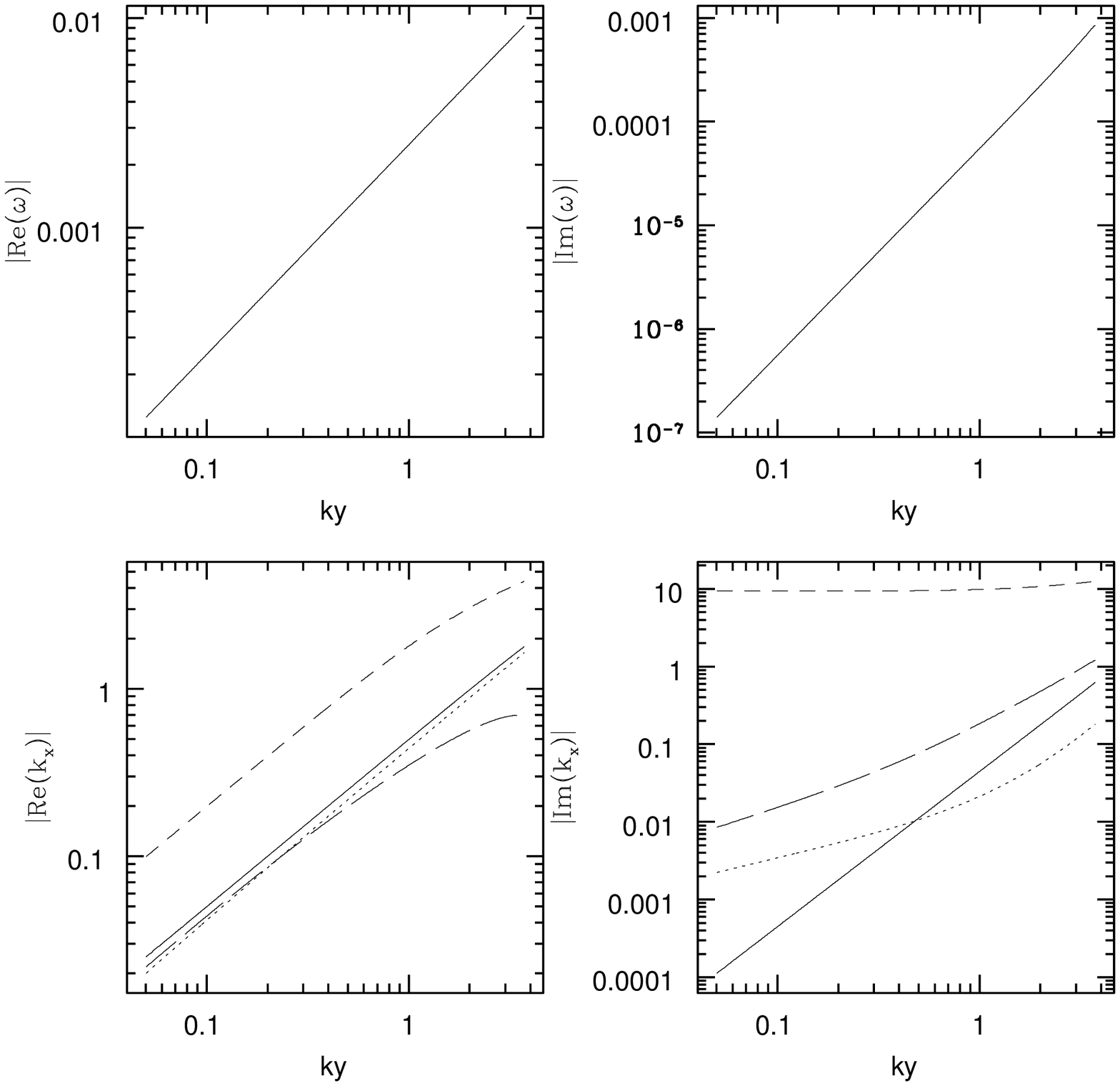}
\caption{Solution of our equations as a function of $k_y$, for the shock structure with accelerated particles' distribution plotted in Fig. \ref{fig:fdistr}, lower panel, dotted line. Please refer to caption of Fig. \ref{fig:onesolution} for details.}
\label{fig:bsolution}
\end{figure}
\begin{figure}
\plotone{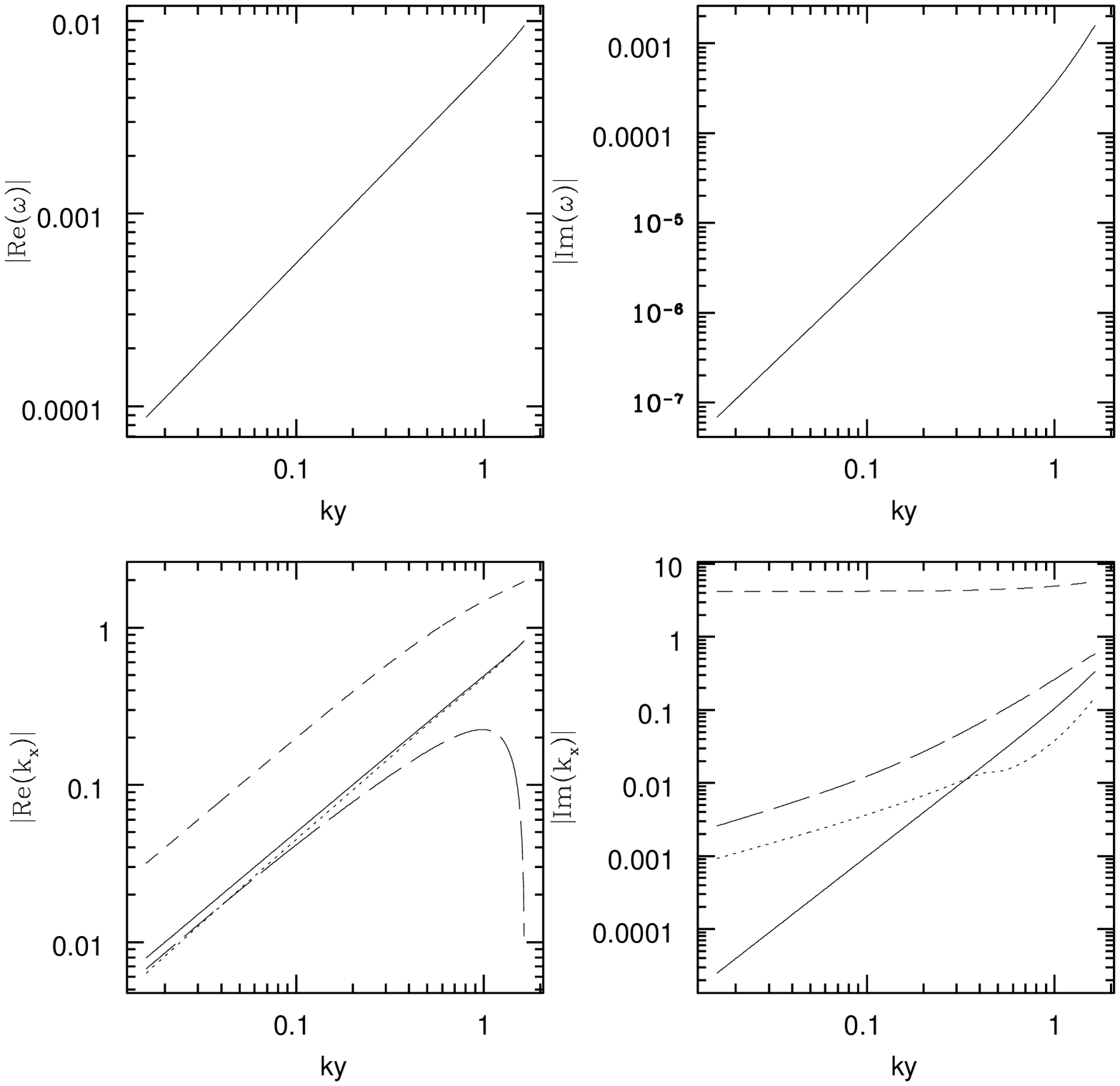}
\caption{Solution of our equations as a function of $k_y$, for the shock structure with accelerated particles' distribution plotted in Fig. \ref{fig:fdistr}, lower panel, dashed line. Please refer to caption of Fig. \ref{fig:onesolution} for details.}
\label{fig:csolution}
\end{figure}

We proceeded in the following way. We calculated the determinant of the system of five eqs. \ref{link1}, \ref{link2}, \ref{link3}, \ref{link4} and \ref{eq:eigenfrequencyl} and we set it to 0, obtaining an equation with 5 unknowns: $\omega$, $k_{xpu}$, $k_{xtu}$, $k_{xpd}$, $k_{xtd}$, which are respectively the frequency and the $x$-components of the wave vectors of the upstream pressure wave,  the upstream third mode, the downstream pressure wave and the downstream third mode. This forms a system of five equations together with four dispersions relations as in eq. \ref{reldisp}, each one linking $\omega$ and $k_y$ with their respective $k_x$. This is exactly what we did above when we solved eqs. \ref{eq:displan} and \ref{eq:eiglan}. In Fig. \ref{fig:onesolution} and in Figs. \ref{fig:asolution}, \ref{fig:bsolution}, \ref{fig:csolution}, we illustrate the solution of our system of equations as a function of $k_y$, respectively for the single and the multiple solution of the shock structure. Absolute values of real parts (left panels) and imaginary parts (right panels) of $\omega$ (upper panels), $k_{xpu}$, $k_{xtu}$, $k_{xpd}$, $k_{xtd}$ (lower panels) are plotted. Signs of these quantities are reported in Tab. \ref{tab:signs}. These solutions must be all discarded because they have some waves diverging at infinity, just like the first four solutions in Tab. \ref{tab:sollan}. We used as starting values in our searching algorithm each one of the six corresponding limit solution in the linear case.  All the solutions we have found cannot be accepted for the same reasons discussed above for the linear regime.

\placetable{tab:signs}
\placefigure{fig:onesolution}
\placefigure{fig:asolution}
\placefigure{fig:bsolution}
\placefigure{fig:csolution}

\begin{table}
\centering
\begin{tabular}{*{5}{c}}
\hline
		& single solution		&			\multicolumn{3}{c}{ multiple solution}					\\
		&Fig. \ref{fig:onesolution}	&Fig. \ref{fig:asolution}	&Fig. \ref{fig:bsolution}	&Fig. \ref{fig:csolution}	\\
\hline	
$\Re(\omega)$	&		+		&		+		&		+		&		+		\\
$\Im(\omega)$	&		-		&		-		&		-		&		-		\\
$\Re(k_{xtu})$	&		+		&		+		&		+		&		+		\\
$\Im(k_{xtu})$	&		-		&		-		&		-		&		-		\\
$\Re(k_{xpu})$	&		+		&		+		&		+		&		+		\\
$\Im(k_{xpu})$	&		-		&		-		&		-		&		-		\\
$\Re(k_{xtd})$	&		-		&		-		&		-		&		-		\\
$\Im(k_{xtd})$	&		+		&		+		&		+		&		+		\\
$\Re(k_{xpd})$	&		-		&		-		&		-		&		-		\\
$\Im(k_{xpd})$	&		-		&		-		&		-		&		-		\\
\hline
\end{tabular}
\caption{Signs of functions plotted in Figs. \ref{fig:onesolution}, \ref{fig:asolution}, \ref{fig:bsolution}, \ref{fig:csolution}. None satisfies eq. \ref{eq:instcond}. No instability has been found by this analysis.}
\label{tab:signs}
\end{table}

\section{Discussion}\label{sec:discussion}

In essence, our method is exact, except for the short-wavelength
(WKBJ) approximation necessary to treat analytically perturbations
in the inhomogeneous upstream precursor. One may however wonder 
where our method differs from previous work \citep{monddrury1998,toptygin1999}, which 
has reported the existence of corrugational instabilities. 

\citet{monddrury1998} have reported the existence of both genuine
corrugational instabilities, and of spontaneous emission of sound
waves, for some (not all) of their models. There is of course a 
number of differences between this paper and theirs: we do not use
the two-fluid approximations, and are interested in small-wavelength
perturbations, contrary to them; we also notice the existence of perturbations
in the upstream fluid, which they do not discuss.  

\citet{toptygin1999} included a number of novelties in his treatment,
but he too did not notice that particles would diffuse upstream, so that
he neglects the third and the d-modes altogether.
Despite this, he does have perturbations upstream (pressure waves!), because he
remarks that, for sufficiently long-wavelength perturbations, these
become supersonic with respect to the fluid alone. This occurs 
because particles and fluid are tightly coupled  in long wavelength
perturbations by diffusion, which traps non-thermal particles, so
that the restoring pressure is the sum of particles' and fluid's
pressures, which is larger than the pure fluid speed. This is of
course correct, and exists in our computations as well.

Contrary to these authors, we have found that shocks with non--vanishing
particle pressure are stable to corrugational instabilities, even 
in the region of parameter space where multiple solutions are possible.
We believe that the reason for this discrepancy lies in our inclusion
of diffusion, and the abandonement of the two-fluid approach, as we now argue. 
We stated in the Introduction that the only possibility for the shock destabilization
(polytropic shocks without particles are well--known to be corrugationally
stable) lies in setting up a loop whereby perturbations shed by the shock
in the downstream region return to the upstream region via particle diffusion
and excite more perturbations; thus perturbation energy is not lost to
downstream infinity, but returns to the shock to create more havoc. However,
the very same mechanism that brings particles back upstream, also causes
perturbations' dampening: we have seen that diffusion leads to damping 
both pressure waves and the so--called third modes. Also, we have established 
that a part of the perturbation occurs in the $d$--mode (where $d$ stands for
damping), where particles are perturbed in phase--space, but the total
perturbation to the pressure vanishes. Thus, some fraction of the 
perturbation goes into a totally useless form (the $d$--mode), the new 
mode (the third mode) is always damped, and even pressure waves acquire a
damping which is altogether neglected in the two--fluid approximation. 
In the end, while diffusion brings particles (and perturbations) back to the 
shock, the simultaneous damping is so strong to make the excitation of an instability
ineffective. 

We should remark that damping of pressure waves is weakest for
the largest wavelengths, a limit which is unaccessible to us because of our
WKBJ approximation, but is exactly the limit investigated by \citet{monddrury1998}.
While we have not found a large wavelength beyond which the shock becomes
unstable, we cannot exclude that a proper treatment of the perturbations 
in the space--dependent precursor may yield a transition to the unstable regime.

A further comment is in order: of all possible geometries, the planar 
one is probably the least likely to display instability. Consider in fact
a spherically symmetric explosion like a SuperNova. In this case, the 
perturbations (except of course for entropy and vorticity) generated 
downstream do not escape to infinity, as they do in the planar case, but 
return to the shock because the downstream region is finite and because
they are deflected by a spatially-dependent refraction index; once they
reach the shock,  they may generate further perturbations. The situation 
is even more promising when the shock is due to an accretion flow, or is 
stalling: in fact, except for the presence of particles, this is exactly 
the scenario proposed (see \citealt{laming2007} for an analytic approach and discussion)
for the generation of asymmetries in proto-neutron stars: in this case,
the mechanism for the instability of a stalled accretion shock is the
reflection by the hard star surface into outgoing pressure waves of 
advected entropy  perturbations, which return to the shock to 
generate more mischief. In the problem with particles, there is the 
extra complication due to diffusion, to be overcome to generate instability.
Given the relative complexity of even a pure fluid analysis \citep{laming2007},
it is likely that this problem will require a numerical approach.

\appendix
\section{D-mode in an homogeneous semi-infinite medium}
\label{app:dmode}

We give here an explicit expression for the d-mode in an homogeneous
but semi-infinite medium. The d-mode is the solution of the
diffusion equation
\begin{equation}
\frac{\partial \delta\! f}{\partial t} = \nabla\cdot(D\nabla\delta\!
f) - u \frac{\partial \delta\! f}{\partial x}\;
\end{equation}
where the speed $u$ is assumed constant because of homogeneity. We
solve first the equation with $u = 0$; to do so, we remark that
suitable boundary conditions are that $\delta\! f\rightarrow 0$ as
$x\rightarrow \pm\infty$, depending on whether we are considering
downstream or upstream regions, respectively. The solution can be
obtained by separation of variables, obtaining:
\begin{equation}
\delta\! f = \alpha(p,k_x) e^{\nu t} e^{k_x x} e^{\pm\imath k_y
y}\;,
\end{equation}
subject to the constraint
\begin{equation}
\nu = D(k_x^2-k_y^2)\;.
\end{equation}
The sign of $k_x$ is the one that allows the solution to remain
finite at infinity. Solutions belonging to different values of $k_x$
and $p$ can obviously be superposed, but we know that there is
another boundary condition (eqs. \ref{decomp1}, \ref{decomp2} and
following discussion) to be satisfied: at the shock, $x = 0$,
\begin{equation}\label{formalgud}
\delta\! f = g(p) e^{\imath\omega t} e^{-i k_y y}
\end{equation}
which obviously gives
\begin{equation}
\nu = \imath\omega\;,\;\;\; Dk_x^2 = \imath\omega + D k_y^2\;,\;\;\;
\alpha(p) = g(p)\;.
\end{equation}
The function $g(p)$ was chosen so that
\begin{equation}
\int p^3 v g(p)\; d\!p = \int D(p) p^3 v g(p)\; d\!p = 0\;.
\end{equation}
We have already seen (see the discussion leading to eq.
\ref{justahelp}) that these are the conditions for the vanishing of
$\delta\! P_c$ and its first time derivative at the initial time,
and thus at all times. The same property is of course acquired by
$\alpha(p)$, so that the mode we just found is surely a d-mode for
the upstream region, in the Homogeneous Approximation.

When we assume $u \neq 0$, the above formulae remain correct except
for the substitution $\nu\rightarrow \nu+u k_x$.

\section{Explicit derivation of functions $w$}
\label{app:funcw}

We need to find the solution to
\begin{eqnarray}
D (k_{xd}-k_{xu})g_u +\frac{u_2-u_1}{3}\frac{\partial
g_u}{\partial\ln p} = &-&\frac{u_2-u_1}{3}A_{iu}\frac{\partial\delta\!
f_{iu}}{\partial\ln p}
-\frac{1}{3}\frac{\partial f}{\partial\ln p}(\delta\! v_{2x}-\delta v_{1x})\nonumber\\
&-&D A_{iu}\delta\!f_{iu}(\imath k_{xiu}+k_{xd}) +D
A_{id}\delta\!f_{id}(\imath k_{xid}+k_{xd})\;.
\end{eqnarray}
This is an equation for $g_u(p)$:
\begin{equation}\label{eqg}
E(p) g_u(p)+F \frac{\partial g_u(p)}{\partial \ln p}=G(p)\;,
\end{equation}
\begin{eqnarray}
E(p)&\equiv& D(p) (k_{xd}(p)-k_{xu}(p))\;,\\
F&\equiv&\frac{u_2-u_1}{3}\;,\\
G(p)&\equiv & -\frac{u_2-u_1}{3}A_{iu}\frac{\partial\delta\!
f_{iu}(p)}{\partial\ln p}
-\frac{1}{3}\frac{\partial f(p)}{\partial\ln p}(\delta\! v_{2x}-\delta\! v_{1x})\nonumber\\
&&\qquad -D(p) A_{iu}\delta\!f_{iu}(p)(\imath k_{xiu}+k_{xd}(p)) +D(p)
A_{id}\delta\!f_{id}(p)(\imath k_{xid}+k_{xd}(p))\;.
\end{eqnarray}
We find a solution of the homogeneous form of eq. \ref{eqg}:
\begin{equation}
E(p) w_C(p)+F \frac{\partial w_C(p)}{\partial \ln p}=0\;,
\end{equation}
\begin{equation}
w_C(p)=\exp\left[-\frac{1}{F}\int_{p_m}^p
{E(p')\frac{dp'}{p'}}\right]
=\exp\left[-\frac{3}{u_2-u_1}
\int_{p_m}^p{(k_{xd}(p')-k_{xu}(p'))D(p')\frac{dp'}{p'}}\right]\;.
\end{equation}
Now we seek a solution of \ref{eqg} of the form $g_u(p)=\tilde{C}(p)
w_C(p)$:
\begin{equation}
E(p)\tilde{C}(p) w_C(p)-\frac{E(p)}{F}F\tilde{C}(p)w_C(p) 
+F w_C(p)\frac{\partial\tilde{C}(p)}{\partial\ln p}=G(p)\;,
\end{equation}
\begin{equation}\label{eqgC}
F w_C(p)\frac{\partial\tilde{C}(p)}{\partial\ln p}=G(p)\;,
\end{equation}
whose solution is:
\begin{equation}\label{solg}
g_u(p)=C
w_C(p)+w_C(p)\frac{1}{F}\int_{p_m}^p{\frac{G(p')}{w_C(p')}\frac{dp'}{p'}}\;.
\end{equation}
Let us define:
\begin{eqnarray}
w_{iu}(p)&\equiv&-w_C(p)\left[
\int_{p_m}^p{\frac{1}{w_C(p')}\frac{\partial\delta\!
f_{iu}(p')}{\partial\ln p'}\frac{dp'}{p'}}\right.\nonumber\\
&&\qquad\left. +\frac{3}{u_2-u_1}\int_{p_m}^p{\frac{(\imath k_{xiu}+k_{xd}(p'))D(p')\delta\!f_{iu}(p')}{w_C(p')}\frac{dp'}{p'}}\right]\;,\\
w_{id}(p)&\equiv &w_C(p)
\frac{3}{u_2-u_1}\int_{p_m}^p{\frac{(\imath k_{xid}+k_{xd}(p'))D(p')\delta\!f_{id}(p')}{w_C(p')}\frac{dp'}{p'}}\;,\\
w_0(p)&\equiv&-w_C(p)\frac{1}{u_2-u_1}
\int_{p_m}^p{\frac{1}{w_C(p')}\frac{\partial f(p')}{\partial\ln p'}\frac{dp'}{p'}}\;,
\end{eqnarray}
therefore we obtain:
\begin{equation}{}
g_u(p)=C w_C(p)+A_{iu}w_{iu}(p)+A_{id}w_{id}(p)+(\delta\!
v_{2x}-\delta\! v_{1x})w_0(p)\;.
\end{equation}

\end{document}